\documentclass[]{article}
\usepackage[margin=1.5in]{geometry}

\usepackage{physics}
\usepackage{float}
\usepackage{physunits}
\usepackage{bm}
\usepackage{upgreek}
\usepackage{amssymb}
\usepackage{graphicx}
\usepackage[font=smaller,labelfont=bf]{caption}
\usepackage{mathtools}
\usepackage{xcolor}
\definecolor{codegreen}{rgb}{0,0.6,0}
\usepackage{hyperref}
\hypersetup{
	colorlinks=true,
	linkcolor=blue,
	citecolor=codegreen
}
\usepackage{pdfpages}
\usepackage{authblk}
\usepackage{dsfont}

\DeclareMathOperator*{\Ass}{\scalerel*{\RR{A}}{\sum}}
\usepackage{scalerel}

\newcommand{\RR}[1]{\mathrm{#1}}
\renewcommand{\eqref}[1]{equation (\ref{#1})}
\newcommand{\pref}[1]{(\ref{#1})}

\newcommand{\Bo}{\mathrm{Bo}}

\newcommand{\Ndf}{{N_\RR{ndof}}}
\newcommand{\Nen}{{N_\RR{en}}}

\newcommand{\vub}{\bm{\upbeta}}
\newcommand{\vuxi}{\bm{\upxi}}

\newcommand{\BB}{\mathcal{B}}

\newcommand{\gen}{\RR{gen.}}

\title{Fluidic Shaping over arbitrary domains: \\theory and high order finite-elements solver}
\author[1]{Amos A. Hari}
\author[1,2]{Moran Bercovici\footnote{Email: mberco@technion.ac.il}}
\affil[1]{Faculty of Mechanical Engineering, Technion - Israel Institute of Technology, Haifa, Israel}
\affil[2]{Department of Materials, ETH Z\"urich, Z\"urich, Switzerland}

\date{}

\begin{document}
\maketitle

\begin{abstract}
	Fluidic Shaping is a novel method for fabrication of optical components based on the equilibrium state of liquid volumes in neutral buoyancy, subjected to geometrical constraints. The underlying physics of this method is described by a highly nonlinear partial differential equation with Dirichlet boundary conditions and an integral constraint.
	To date, useful solutions for such optical liquid surfaces could be obtained analytically only for the linearized equations and only on circular or elliptical domains. A numerical solution for the non-linear equation was suggested, but only for the axi-symmetric case. 
	Such solutions are, however, insufficient as they do not capture the full range of optical surfaces. Arbitrary domains offer an important degree of freedom for creating complex optical surfaces, and the nonlinear terms are essential for high quality solutions. Moreover, in the context of optics, it is not sufficient to resolve the shape of the surface, and it is essential to obtain accurate solutions for its curvature, which governs its optical properties.

	We here present the theoretical foundation for the Fluidic Shaping method over arbitrary domains, and the development of a high order (quintic) finite element numerical solver, capable of accurately resolving the topography and curvature of liquid interfaces on arbitrary domains.
	The code is based on reduced quintic finite elements, which we have modified to capture curved boundaries.
	We compare the results against low order finite elements and non-deformed high order elements, demonstrating the importance of high order approximations of both the solution and the domain.
	We also show the usability of the code for the prediction of optical surfaces derived from complex boundary conditions.
\end{abstract}

\section{Introduction}
Fluidic Shaping is a novel fabrication method for creating optical components originally introduced by Frumkin and Bercovici in 2021.\cite{FrumkinFlow}
In this method, a volume of liquid polymer is submerged in an immiscible immersion liquid of equal density. This induces neutral buoyancy on the liquid polymer, and by pinning it to a geometrical boundary condition it is possible to obtain a desired optical component shape.
Unlike traditional optical manufacturing methods, the Fluidic Shaping approach does not require any mechanical grinding or polishing, and yet it produces components of sub-nanometric surface quality thanks to the natural smoothness of liquid interfaces.
In their work, Frumkin and Bercovici introduced the principle concepts of the method and provided a linearized analytical solution for axi-symmetric lenses obtained using a circular bounding frame as the geometrical constraint.
Elgarisi et al.\cite{ElgarisiOptica} expanded the use of Fluidic Shaping for rapid fabrication of freeform optical components.
They generalized the mathematical model and allowed for height variations along the circular bounding frame.
Through an analytical solution of the linearized model and its experimental validations, they characterized the range of freeform surfaces that can be produced using Fluidic Shaping over circular domains.
In 2024, Na et al.\cite{naKaust2024} introduced an optimization methodology for axi-symmetric circular lenses fabricated by Fluidic Shaping.
Unlike previous works, which provided approximated analytical solutions, Na et al. numerically solved the one-dimensional nonlinear axi-symmetric equation. They showed that this representation is more accurate than the approximated analytical solutions, especially in cases where the assumptions of linearization are no longer valid.
They implemented the numerical solution in a ray tracing software and used it to simulate imaging through a Fluidic Shaping lens; then, using an optimization technique they designed the lens to provide the best image.

In 2025, Elgarisi et al.\cite{elgarisi2025} explored whether Fluidic Shaping is a viable manufacturing approach for common ophthalmic lenses.
They showed that by deviating from a circular bounding frame to an elliptical one, ophthalmic lenses with both astigmatic and spherical corrections can be created.
Hence, we conclude that variations in the boundary footprint introduce a useful degree of freedom for lens design, with which one may satisfy a wide range of optical prescriptions.
This conclusion motivates us to expand the Fluidic Shaping theory to account for arbitrary domains.

On an arbitrary domain, the task of solving a highly nonlinear boundary value problem must be approached numerically.
The specific requirements for such numerical solver depend on the application.
In our optical context the solution should at the very least provide accurate information about the unknown surface and its first and second derivatives. This information is required to calculate the curvature of the surface, which determines the optical power of the corresponding lens.
For precision optic applications, where the shape of the lens must be accurate to a fraction of the operation wavelength, particular accuracy is necessary.
For example, a $1\RR{cm}$ diameter lens operating in the visible range ($\lambda \simeq 500\RR{nm}$) requires shape accuracy on the order of $\lambda/10=50\RR{nm}$, i.e., $6$ orders of magnitude smaller than the characteristic geometrical length.

There exist a multiplicity of works dedicated to numerically finding the position of a static liquid interface.\cite{walzelPerturbingCatenoidStability2022,prabhalaPerturbationSolutionShape2010,brandonSimulatedContactAngle1997,zhuShapeForceAnalysis2015,gongLiquidBridgeMorphology2025,dunlopIdentitiesDropletsCircular2020,liuJurinsLawRevisited2018,timmSecondorderaccurateApproximationShape2023}
However, the approaches described in these works are either low order or tailored for specific geometries, and thus not suitable for optical applications.
The most notable solver in the field is Brakke's Surface Evolver \cite{SurfaceEvolver}, which solves for minimum energy surfaces with nearly arbitrary three-dimensional geometrical constraints.
Among its many features, Surface Evolver allows for high order Lagrangian discretization of the solution surface, thus making it an excellent candidate for accurate surface computations, as needed for optical applications. However, even for high order discretization, Surface Evolver does not provide second derivatives information as part of its output, hence limiting its appeal to our needs.

To fully explore the degrees of freedom offered by Fluidic Shaping, and in the absence of suitable tools that can provide the necessary accuracy for optical applications, we turn to developing a dedicated solver.
We here present the theory and implementation of a high order (reduced quintic) finite elements procedure to solve the nonlinear boundary value problem of Fluidic Shaping over arbitrary domains.
We take special care to reduce the geometry mismatch error arising from the spatial discretization, and later, in Section \ref{sec:results}, we show the importance of doing so.
We verify the results of our code against both manufactured and known analytical solutions, and show excellent convergence of the results.
We compare the solutions of the nonlinear equation against linearized analytical solutions which highlight the importance of the nonlinear terms.
Furthermore, we show how our code can be used to assist in designing Fluidic Shaping based lenses; using our code we can predict the nominal optical properties of the lens, and provide insights on the effects of manufacturing simplifications and production inaccuracies (e.g., frame defects, or volume inaccuracies) on the final results.
Lastly, we show the use of our code for designing and analyzing different configuration of micro-lens arrays made using the Fluidic Shaping method.

\section{Problem Statement}\label{sec:problem-statement}
\begin{figure}
	\centering
    \includegraphics[width=\linewidth]{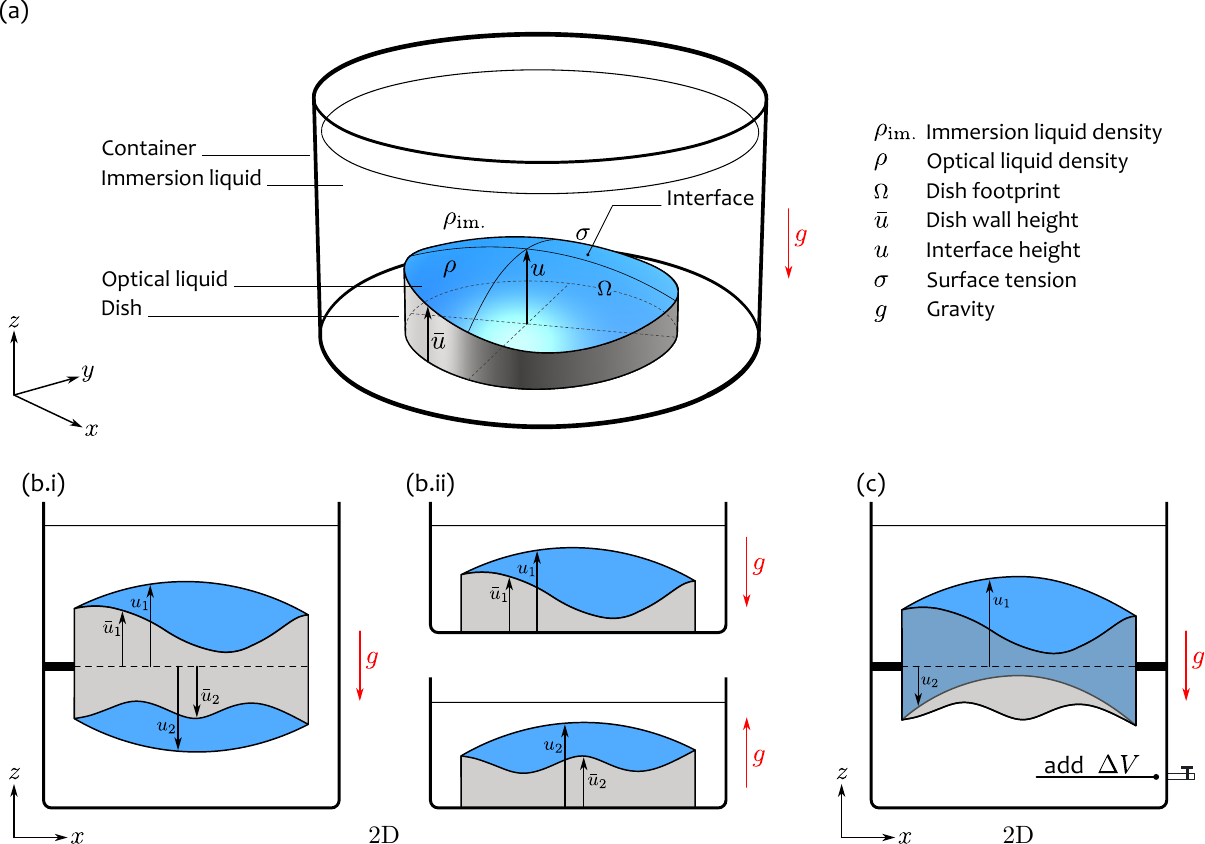}
	\caption{Illustration of the Fluidic Shaping configurations that can be described as a minimization of \eqref{eq:energy} subjected to the volume constraint \eqref{eq:constraint}.
        (a) The basic examined configuration. We submerge a dish, filled with optical liquid of density $\rho$, in a container, filled with immersion liquid of density $\rho_\RR{im}$.
        We assume that the optical liquid is perfectly pinned at height $\bar{u}$ on the dish's edge, and that the shape of the interface $u$ is determined by a balance of surface tension $\sigma$, gravity $g$ and the total optical liquid volume $V$.
		(b) A multiple interface configuration. For each free surface we imagine a corresponding dish as in the basic configuration. The optical liquid volumes is shared between all dishes, and the direction of gravity may flip with the dish's orientation.
		(c) A multiple interface configuration with an additional volume constraint. We take the multiple surface configuration and partition the immersion container to create a sealed closure around the bottom surface. By adding additional immersion liquid volume $\Delta V$ to the sealed closure, we introduce residual pressure which displaces the interfaces.
        }
	\label{fig:system}
\end{figure}

Figure \ref{fig:system}(a) presents the simplest Fluidic Shaping configuration that will serve as the basis of our analysis, and from which more complex configurations will be derived.
Consider a dish with a flat bottom, an arbitrary footprint and a non-uniform wall height.
This dish is filled with an optical liquid of density $\rho$, which we assume to be perfectly pinned to the edge of the wall.
The entire configuration is submerged in a container full of immiscible immersion liquid of density $\rho_\RR{im}$, which introduces a buoyancy force that acts against gravity.
We wish to find the shape of the optical liquid at rest.

Let $l_c$ be the characteristic length of the system, which can be conveniently defined as the diameter of the smallest circle that inscribes the dish's bottom surface.
We define $\Omega\subset\mathbb{R}^2$ as the dimensionless domain of definition which corresponds to the dish's bottom surface, scaled by $l_c$ in its two dimensions, and $\Gamma = \partial\Omega$ as the boundary of this domain.
We define $u:\Omega\rightarrow\mathbb{R}$ to be the function representing the height of the liquid-liquid interface, and the pinning condition is thus defined as $u(x,y) = \bar{u}(x,y)$, where $\bar{u}:\Gamma\rightarrow\mathbb{R}$ is a function that describes the height of the dish's wall along the perimeter.
At steady state, the liquid takes a shape that minimizes the total energy of the system, which in its dimensionless form (scaled by $\sigma l_c^2$) is
\begin{equation}\label{eq:energy}
	\mathcal{E}[u] = \int_{\Omega}\qty[\sqrt{1 + u_{,x}^2 + u_{,y}^2} - \frac{1}{2}\Bo u^2] \dd{\Omega},
\end{equation}
subjected to the global volume constrain (scaled by $l_c^3$),
\begin{equation}\label{eq:constraint}
	 \int_{\Omega} u \dd{\Omega} = V.
\end{equation}
Using the method of Lagrange multipliers, we seek $u$ and $P$ for which the variation of the following functional
\begin{equation}\label{eq:repr-prob}
	\Pi[u;P] = \int_{\Omega}\qty[\sqrt{1 + u_{,x}^2 + u_{,y}^2} - \frac{1}{2}\Bo u^2] \dd{\Omega} - P \qty(V - \int_{\Omega} u \dd{\Omega}),
\end{equation}
vanishes.
The first term in the first integral represents the surface tension energy, which is proportional to the area of the interface.
The second term in that integral represents the gravitational potential energy, with $\Bo = (\rho_{\RR{im}}-\rho)gl_c^2/\sigma$.
This Bond number dictates the relative importance of gravitational force to surface tension, where for $\Bo\gg1$ the system is dominated by gravity, whereas for $\Bo\ll1$ surface tension dominates.
The last term in the functional corresponds to the volume constraint, where $P$ is the Lagrange multiplier.
We thus seek $u\in S$ and $P\in\mathbb{R}$ that yield a stationary value for this functional, where $S = \{u\big|u\in H^{1}(\Omega),\ u=\bar{u}\ \RR{in}\ \Gamma\}$ and the Sobolev space $H^1(\Omega)$ is the set of all square integrable functions with square integrable derivatives over $\Omega$.

Adopting this description, more complex configurations can also be analyzed. Figure \ref{fig:system}(b.i) illustrates a configuration where instead of a dish we use a frame, with the liquid pinned to both its top and bottom edges, i.e., there are now two interfaces to solve for.
The solution to this problem follows directly from the solution to the previous one:
imagine splitting the frame into two separate dishes as depicted in Figure \ref{fig:system}(b.ii).
The first dish has a footprint $\Omega_1$, wall height $\bar{u}_1$, interface height $u_1$ and experiences a gravity of $-g\vb{e}_z$. The second dish has a footprint $\Omega_2$, wall height $\bar{u}_2$, interface height $u_2$, and in its rotated frame of references it experiences a gravity $+g\vb{e}_z$.
While these are seemingly separate problems, there is only one volumetric constraint that applies jointly to both domains. Namely, the shared volume between the dishes is $V$.
To consider both domains simultaneously, we redefine the Bond number as a composite functions, $\Bo(x,y):\Omega\rightarrow\mathbb{R}$. Here $\Omega = \Omega_1\cup\Omega_2$, and $\Bo(x,y) = \Bo$ for $(x,y)\in\Omega_1$ while $\Bo(x,y) = -\Bo$ for $(x,y)\in\Omega_2$.
Similarly, we define $u$ as a composite function of $u_1$ over $\Omega_1$ and $u_2$ over $\Omega_2$, and likewise for $\bar{u}$.
As a result, we may simply replace $\Bo$ with $\Bo(x,y)$ in \eqref{eq:repr-prob}, and plug into it the composite $\Omega$, $u$ and $\bar{u}$.
Thus, any technique we develop for the basic problem of Figure \ref{fig:system}(a) may also be applied for configurations with multiple interfaces, such as that of Figure \ref{fig:system}(b.i).

Another configuration, which is of particular interest to eyewear applications, is that of a meniscus lens where the two optical surfaces are curved in the same direction.
To achieve this using Fluidic Shaping, the immersion liquid container is separated into two parts forming a sealed closure around the bottom part of the lens, as depicted in Figure \ref{fig:system}(c).
When injecting additional immersion liquid, $\Delta V$, into this closure, the optical liquid is deformed to accommodate for this change in volume.\cite{FrumkinFlow,elgarisi2025}
To solve for the shape of the interfaces in this configuration, we first solve the non-inflated case as in Figure \ref{fig:system}(b.i), and calculate the partial volumes $V_1$ and $V_2$ that reside in each of the virtual dishes.
We then decouple the two domains, and solve for $u_1$ over $\Omega_1$ using the volume constraint $V_1 + \Delta V$, and for $u_2$ over $\Omega_2$ using the volume constraint $V_2 - \Delta V$.

In what follows we thus focus on the solution for a dish with a flat bottom of arbitrary shape, arbitrary wall height, and variable Bond number.
Using this method, the solution of any of the problems of Figure \ref{fig:system}(a-c) as well as expansion to other cases, can be readily obtained.

\section{Solution Method}\label{sec:solution-method}
\begin{figure}
	\centering
    \includegraphics[width=\linewidth]{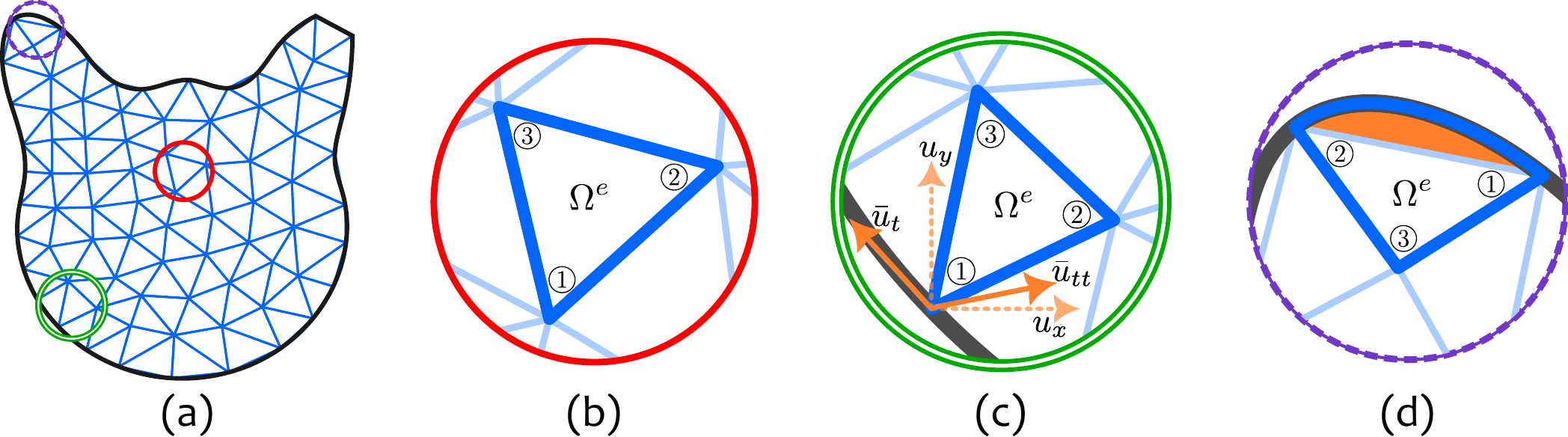}
    \caption{
        Treatment of internal and boundary elements in a triangular mesh of an arbitrary domain.
        (a)
        The triangular discretization of a general domain gives rise to three types of elements:
        (b)
        An element in the bulk of the mesh with no overlap with the boundary.
        This is the most common element in the mesh and it doesn't warrant any special treatment.
        (c)
        An element with one node on the boundary.
        The additional degrees of freedom of the high order element do not naturally coincide with the prescribed boundary condition, requiring a local transformation.
        (d)
        An element with an edge on the boundary.
        In addition to transforming the nodal degrees of freedom, the edge itself should be deformed to maintain a high order accuracy.
    }
	\label{fig:mesh}
\end{figure}
The goal of this work is, for any arbitrary domain $\Omega$, to find the function $u$ and number $P$ for which the functional $\Pi[u,P]$, defined in \eqref{eq:repr-prob}, obtains a stationary value.
To this end, we adopt the finite elements method as the solution approach and, following the standard Galerkin-Ritz technique, replace the unknown function $u\in S$ in \eqref{eq:repr-prob} with $u^h\in S^h$, where $S^h$ is a finite element approximation space.
This, in turn, transforms the functional $\Pi[u,P]$ into a function $\Pi^h(\vb{d},P)$, where $\vb{d}$ is the finite elements displacement vector.\cite{hughes,bathe}

In the context of Fluidic Shaping of optical components, a precise solution for the liquid interface that serves as the optical surface is required.
Moreover, such a solution must accurately approximate the first and second order derivatives of the surface, as these are needed to quantify key optical properties, such as spherical power and astigmatism.
In light of these requirements, we adopt the concept proposed by Bell\cite{bell1969} for reduced quintic triangular finite elements, as they are known to span $S^h$, the approximation space of $S$, with high order accuracy and continuity.\cite{cowper1969,strangAnalysisFiniteElement2008}

Consider the arbitrary domain $\Omega$ which, as depicted in Figure \ref{fig:mesh}(a), we triangulate into multiple subdomains $\Omega^e$, where $e$ denotes the element number.
For those elements that are fully contained within the domain, such as in Figure \ref{fig:mesh}(b), the finite elements approximation is given by
\begin{equation}\label{eq:solution_in_element}
    (u^h)^e = \sum_{a=1}^{\Nen}\sum_{r=1}^{\Ndf} \phi_{(ar)}^e d_{(ar)}^e,
\end{equation}
with $\Nen=3$ as the number of element nodes, and $\Ndf=6$ as the number of nodal degrees of freedom.
Here, $d_{(ar)}^e$ are the six nodal degrees of freedom which correspond to the solution values and derivatives ($u$, $u_x$, $u_y$, $u_{xx}$, $u_{yy}$, and $u_{xy}$) at the respective node, and $\phi_{(ar)}^e:\Omega^e\rightarrow\mathbb{R}$ are the quintic polynomial shape functions that correspond to degree of freedom $r$ in node $a$ of element $e$.

To construct the reduced quintic shape functions, we define
\begin{equation}\label{eq:phi_dof_der}
    \qty{\phi_{(ar),s}}_{s=1}^\Ndf:=\qty{\phi_{(ar)}, \phi_{(ar),x}, \phi_{(ar),y}, \phi_{(ar),xx}, \phi_{(ar),yy}, \phi_{(ar),xy}},
\end{equation}
and require that
\begin{equation}\label{eq:condition_1}
    \phi_{(ar),s}^e(x_b, y_b) = \delta_{(ar)(bs)} = \begin{dcases} 1, &a=b\ \RR{and}\ r=s\\ 0, &\RR{else}\end{dcases},
\end{equation}
where $(x_b,y_b)$ is the position of node number $b$ of the element.
The system of equations \pref{eq:condition_1} provides $18$ of the $21$ constrains needed to fully determine the coefficients of the element shape functions.
For the reduced quintic elements, the final $3$ coefficients are determined by requiring the normal derivatives along each edge to vary as a cubic polynomial.\cite{bell1969,cowper1969,strangAnalysisFiniteElement2008}
Appendix \ref{app:element_formulation} provides the technical details for our implementation of the shape functions.

In Figure \ref{fig:mesh}(c) we consider an element with one node on the boundary $\Gamma$.
From the pinning (Dirichlet) boundary condition, we know that the solution on the boundary is given by $u=\bar{u}$ on $\Gamma$.
However, the function $\bar{u}$ itself is defined solely on the boundary, usually through a parameterization of the boundary curve, i.e., $\bar{u}(\bar{x}(t),\bar{y}(t))$; as a result, we need to express the nodal degrees of freedom ($u$, $u_{x}$, $u_{y}$, $u_{xx}$, $u_{yy}$, and $u_{xy}$) in terms of the boundary data ($\bar{u}$, $\bar{u}_t$, and $\bar{u}_{tt}$).
We adopt the idea by Wu et al.\cite{wuC1ConformingQuadrilateral2020}, who addressed this problem by introducing a local transformation $\vb{T}$ on the vector of nodal degrees of freedom $\vb{d}_a^e=\qty{d_{(ar)}^e}_{r=1}^{\Ndf}$, such that after the transformation, the components of the new vector, denoted $\vb{d}_a^{e*} = \vb{T}\vb{d}_a^e$, directly correspond to values of $\bar{u}$ and its derivatives.
This transformation takes a different form, depending if $\bar{u}$ is smooth or not, and in Appendix \ref{app:dof_trans} we provide the derivation of the explicit transformation for both the smooth and sharp point cases.
Thus, for elements with nodes on the boundary, such as depicted in Figure \ref{fig:mesh}(c), the finite elements approximation can be expressed by \eqref{eq:solution_in_element} with $d_{(ar)}^e$ and $\phi_{(ar)}^e$ replaced respectively with
\begin{equation}\label{eq:dof_transform}
    d_{(ar)}^{e*} = \sum_{c=1}^{\Nen}\sum_{t=1}^{\Ndf}T_{(ar)(ct)}d_{(ct)}^{e} \qq{and} \phi_{(ar)}^{e*} = \sum_{b=1}^{\Nen}\sum_{s=1}^{\Ndf}\phi_{(bs)}^{e}Q_{(bs)(ar)},
\end{equation}
where $\vb{Q}=\vb{T}^{-1}$.
Note that this representation can be used for all nodes, where for an internal nodes $\vb{T}$ is just the $\Ndf\times\Ndf$ identity matrix.

Lastly, in Figure \ref{fig:mesh}(d) we show the case where a full edge of the element is on the boundary.
Each of the nodal degrees of freedom can be treated using the above transformation.
However, as the Figure illustrates, the triangulation of the domain may misrepresent the shape of the boundary wherever it is curved.
As later shown in Figure \ref{fig:convergence}, the error introduced by this geometrical mismatch may be significant enough to reduce the order of accuracy 
down to $\order{h^2}$, with $h$ being the \emph{mesh parameter}\cite{hughes} of the finite elements mesh.
To remedy this, a common practice is to replace the triangular element with a deformed element whose edge matches the boundary.\cite{ciarletInterpolationTheoryCurved1972}
In appendix \ref{app:element_formulation} we provide a detailed derivation for such a transformation. 
We note that this transformation does not modify the nodal degrees of freedom but rather manifests itself through the Jacobian of the finite element integrals.
In Appendix \ref{app:galerkin_ritz_fe_formulation} we provide the full details and formulas for the Galerkin-Ritz finite element formulation that account for such boundary transformations.

\section{Results}\label{sec:results}
\begin{figure}
	\centering
	\includegraphics[width=\linewidth]{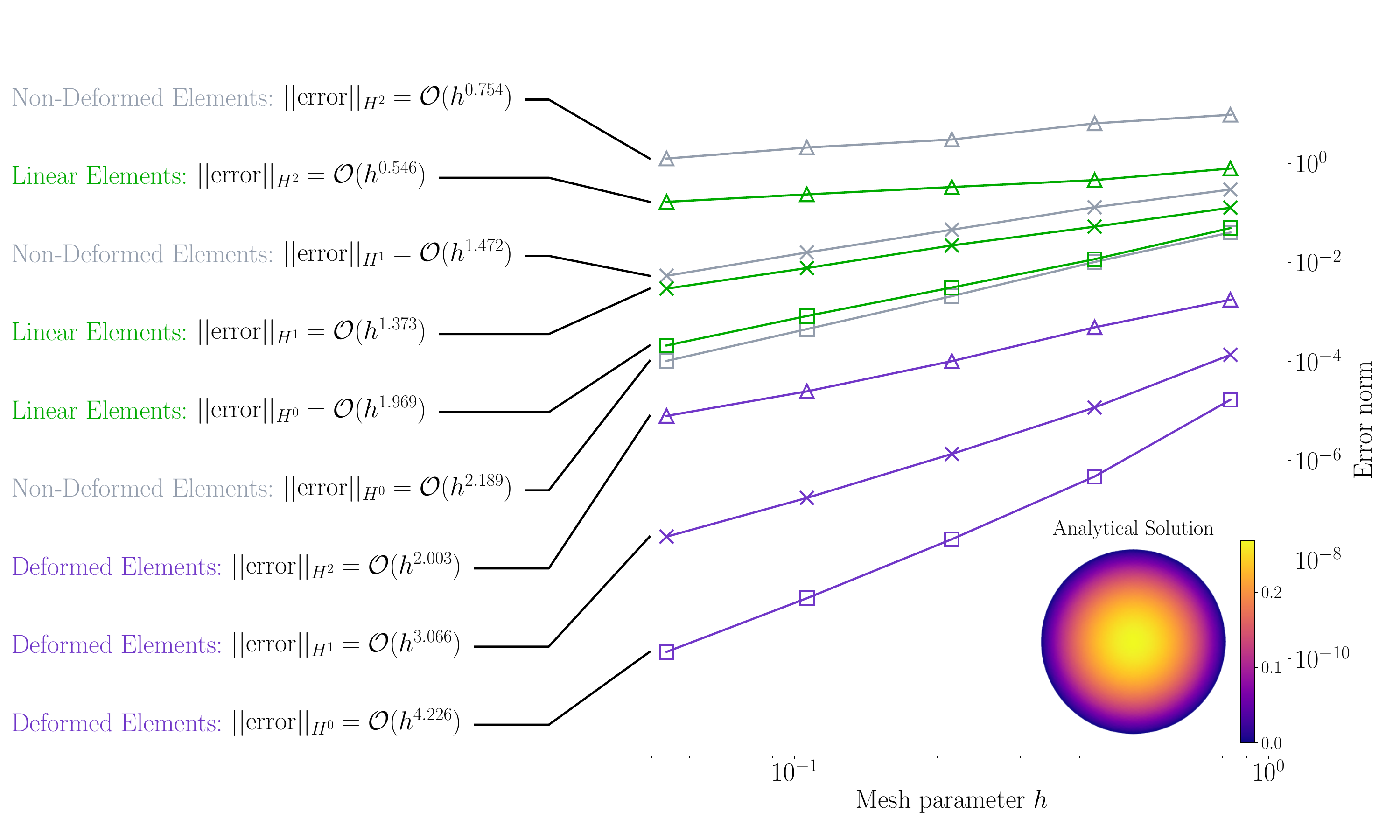}
    \caption{
        Accuracy analysis of our deformed elements method.
        The implementation of our deformed finite elements method (purple) improves solution accuracy compared with classical approaches: linear elements (green) and non-deformed reduced quintic elements (grey).
        We solve the Fluidic Shaping problem for a spherical lens over a circular domain, compare each numerical solution with the analytical spherical surface, and report the errors and apparent convergence rate for each method.
        Across the tested meshes, our deformed elements method outperforms the others by orders of magnitude, and its convergence rate suggest even larger gains for finer discretizations.
	The $H^0$ norm is defined as $\norm{\RR{error}}_{H^0} := \sqrt{\int_{\Omega} (u-u_\RR{exact})^2 \dd{\Omega}}$, the $H^1$ semi-norm as $\norm{u}_{H^1} := \sqrt{\int_\Omega (u-u_\RR{exact})_{,i}(u-u_\RR{exact})_{,i} \dd{\Omega}}$, and the $H^2$ semi-norm as $\norm{u}_{H^2} := \sqrt{\int_{\Omega} (u-u_\RR{exact})_{,ij}(u-u_\RR{exact})_{,ij} \dd{\Omega}}$.
        To obtain high order derivative for the linear finite element solution, we used the Clough-Tocher interpolation, which yields a continuous differentiable interpolant which approximately minimize surface curvature.\cite{CloughTocher2DInterpolatorSciPyV1162}
        }
	\label{fig:convergence}
\end{figure}
Using the method described in Section \ref{sec:solution-method}, and the formulas provided in Appendix \ref{app:galerkin_ritz_fe_formulation}, we implemented a finite elements solver for the Fluidic Shaping problem over arbitrary domains.
To verify the correctness of our solver, we force it to yield a known result using the method of manufactured solutions \cite{roacheMMS2002}, and we compare the manufactured numerical solution to the analytical one.
Specifically, we prescribed a spherical solution over a circular  domain, and in Figure \ref{fig:convergence} we present several norms of the error as a function of the mesh parameter.
To compare, the error norms of high-order non-deformed elements, and of linear elements, are also presented.
The deformed elements outperform the non-deformed and linear ones, providing an order of accuracy of $\order{h^{4.226}}$ vs $\order{h^{2.189}}$ (non-deformed) and $\order{h^{1.969}}$ (linear) for the $H^0$ norm, $ \order{h^{3.066}} $ vs $\order{h^{1.472}}$ and $ \order{h^{1.373}} $  for the $H^1$ norm, and $\order{h^{2.003}}$ vs $\order{h^{0.754}}$ and $\order{h^{0.546}}$ for the $H^2$ norm.
Interestingly, the transition from linear elements to high-order non-deformed ones provides only marginal improvement in the order of accuracy, whereas the transition to deformed elements provides a major increase.
These results show the predominance of the geometry mismatch error, which is introduced when the domain $\Omega$ in \eqref{eq:repr-prob} is replaced by $\Omega^h$.
Essentially, the source of this error is the Boolean symmetric difference of the domains,  $\Omega\Delta\Omega^h$.
When $\Omega^h$ is deformed to better match the shape of $\Omega$, this error decreases in magnitude and its rate of  convergence improves.
Deforming the boundary elements is thus essential to preserve the accuracy of the high order finite elements when solving over arbitrary domains.

\begin{figure}
	\centering
	\includegraphics[width=\linewidth]{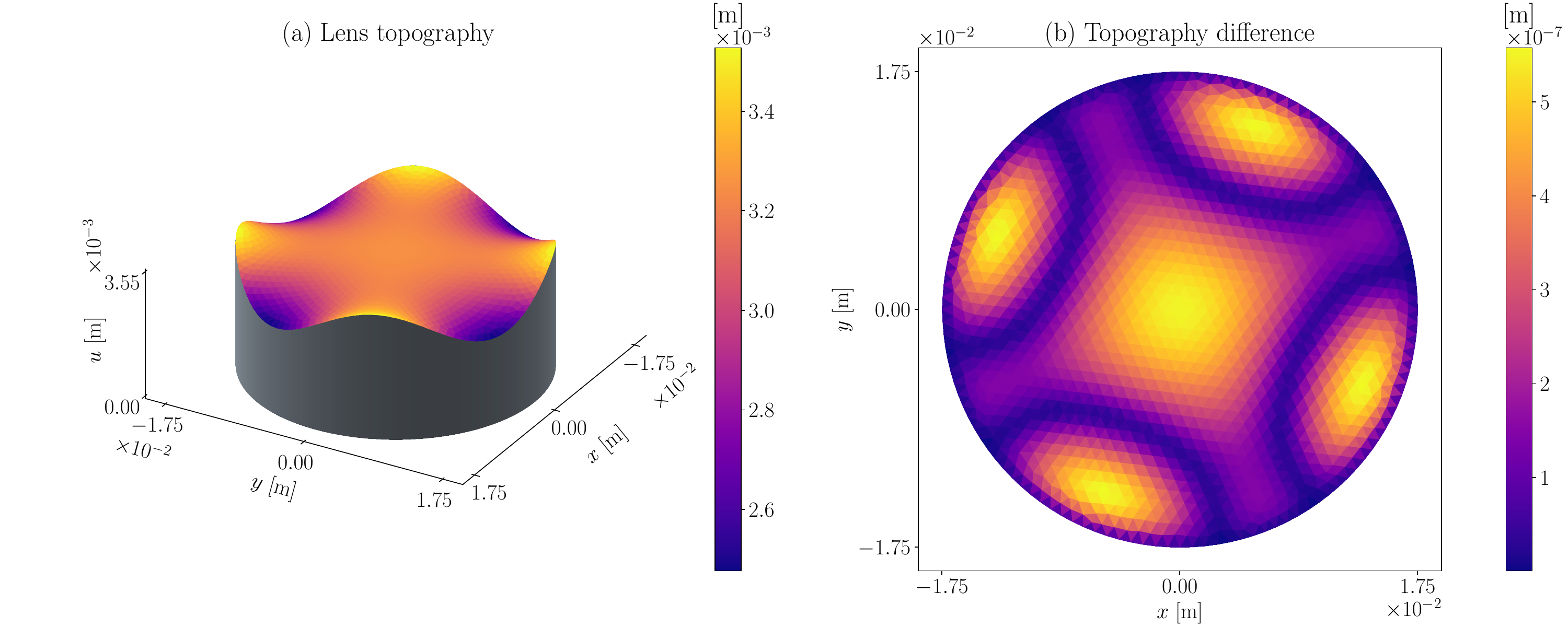}
    \caption{
        High precision optics demand nonlinear terms.
        Elgarisi et al.\cite{ElgarisiOptica} showed that the shape of a circular lens can be approximated analytically through linearization of the Fluidic Shaping problem, \eqref{eq:repr-prob}.
        However, by comparing the numerical results of our method to the analytical results of Elgarisi et al. we show that the linear analytical solution is not accurate enough for high precision optics.
        (a) An isometric view of the lens as obtained from the numerical solver.
        (b) The absolute difference in the topography between the analytic and numeric surfaces, showing deviations of 500 nm - beyond what is allowed for precision optics. 
        This lens was obtained for diameter $3.5$ cm, volume $3$ ml, Bond $3$, and height $3 + 0.55\sin(4\theta)$ mm, for $\theta\in[0,2\pi)$.
    }
	\label{fig:elgarisi}
\end{figure}
With the solver verified, we may now proceed to demonstrate its value for various optical applications.
As mentioned in the introduction, precision optical elements require an accuracy on the order of $\lambda/10$, i.e., approximately $50$ nm for wavelengths in the visible range (for which $\lambda\simeq500$ nm).
To date, the only available solution for Fluidic Shaping with variable boundary conditions is the analytical one by Elgarisi et al. \cite{ElgarisiOptica} for a circular domain.
However, that solution was obtained assuming small deformations of the optical surfaces, which allowed to linearize the governing equation.
Thus, it is not clear whether this analytical solution is accurate enough for precision optics.
In Figure \ref{fig:elgarisi}, we examine the $3.5$ cm-diameter freeform lens presented by Elgarisi et al., which is created by superposing a $550$ $\RR{\mu m}$ sinusoidal variation on a $3$ mm high  boundary wall.  We compare the linear analytical solution with the numerical one, which we solve  using a mesh parameter of $h \simeq 1.242\times10^{-3}$ m, corresponding to $h \simeq 7.102\times10^{-2}$ in the dimensionless units of Figure \ref{fig:convergence}. Under these conditions, we expect a numerical error on the order of magnitude of $10^{-11}$ m for the deformed elements.
The results show differences in topography that reach approximately $500$ nm, an impressive feat for the linear analytical model, but one which is an order of magnitude too large for precision applications.
Moreover, the numerical solution can also be used to determine the region of parameters for which the analytical solution is sufficiently accurate. 
For example, in Appendix \ref{app:small_boundary_variations} we present a similar optical component with a boundary variation of $100$ $\RR{\mu m}$ (instead of $550$ $\RR{\mu m}$) that better satisfies the linearization approximation of $u_{x}, u_{y}\ll1$ in the Elgarisi model.
There, the difference between the solutions drops to $25$ nm, suggesting that the analytical solution can be safely used.

\begin{figure}
	\centering
	\includegraphics[width=\linewidth]{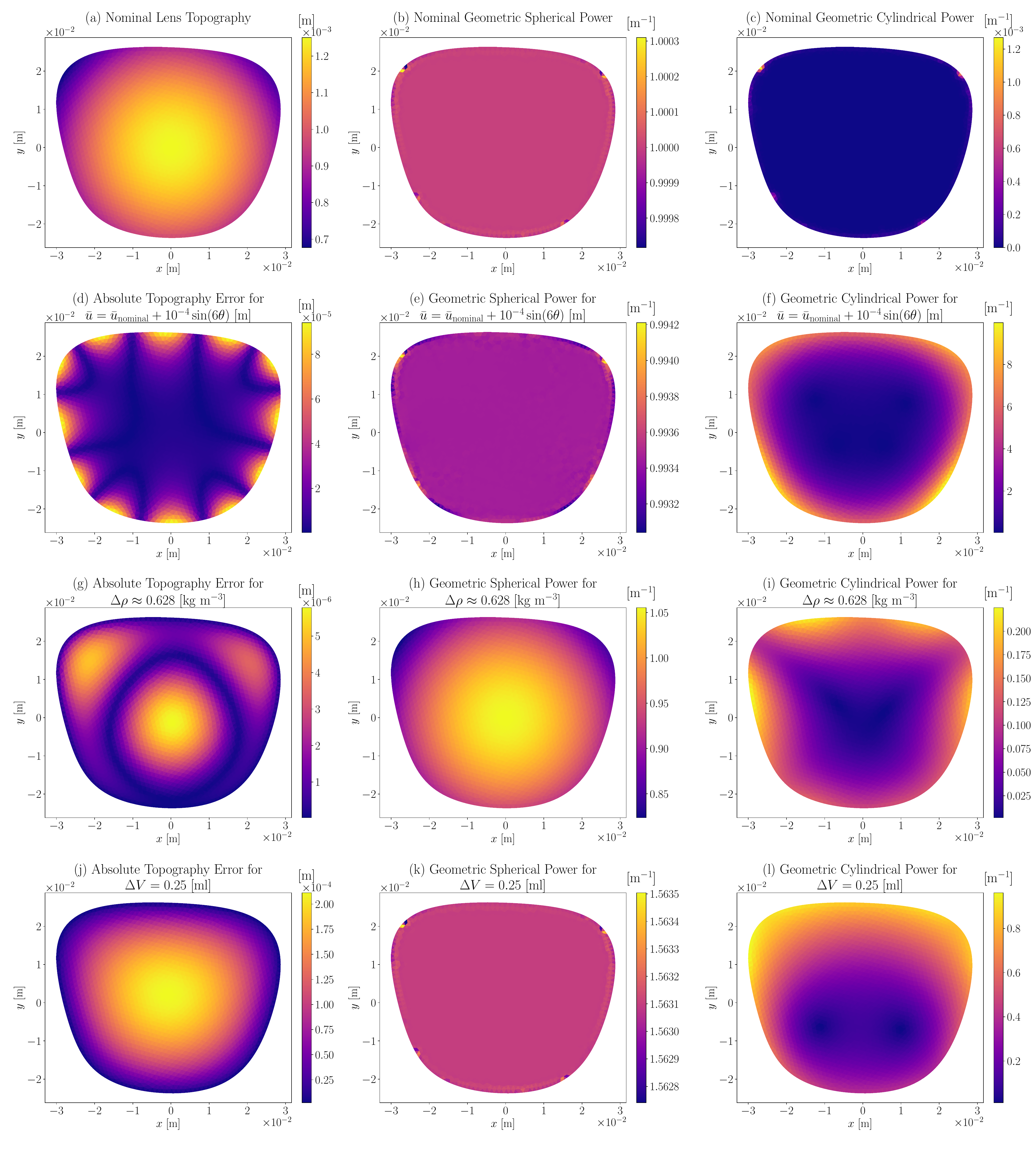}
    \caption{Sensitivity analysis of a spherical ophthalmic lens manufactured using the Fluidic Shaping method on an Oakley Slender Satin eyewear frame.
	(a-c) The computed nominal topography, geometric spherical power, and geometric cylindrical power.
	We investigate the process' sensitivity to errors in (d-f) the boundary frame height, (g-i) the immersion liquid density (which alters the Bond number to  $\Bo=2$ for surface tension of $7.7$ $\RR{mN/m}$ \cite{naKaust2024}), and (j-l) the injection of excess lens liquid. 
    The geometric spherical power is defined as $-(\kappa_x+\kappa_y)/2$ and the geometric cylindrical power  as $(\kappa_x-\kappa_y)$, where $\kappa_x$ and $\kappa_y$ are the surface curvature in the $x$ and $y$ directions.}
	\label{fig:input_error}
\end{figure}
Next, we consider Fluidic Shaping of ophthalmic eyewear lenses.
With our solver, we are able to compute, for the first time, the Fluidic Shaping problem over arbitrary domains. 
Figure \ref{fig:input_error}(a) shows the spherical surface topography obtained by Fluidic Shaping on an eyewear-shaped domain.
Figures \ref{fig:input_error}(b) and \ref{fig:input_error}(c), respectively, show the geometric spherical power, $ -(\kappa_x+\kappa_y)/2 $, and geometric cylindrical power, $ (\kappa_x - \kappa_y) $, of the lens, where $\kappa_x$ and $\kappa_y$ are the surface curvature in the $x$ and $y$ directions.
Because the lens surface is nominally spherical, then its geometric spherical power and cylindrical powers are, respectively, constant and zero.
Numerical errors in the calculation of the derivatives lead to deviations close to the boundary of the domain, but their magnitude is negligible in the context of ophthalmic applications (4th digit in spherical power).
We can now use our solver to study the sensitivity of the Fluidic Shaping process to typical manufacturing errors.
For example, we investigate how the solution changes for either deviations in the boundary shape (Figures \ref{fig:input_error}(d-f)), errors in immersion liquid density (Figures \ref{fig:input_error}(g-i)), or injection of excess lens volume (Figures \ref{fig:input_error}(j-l)).
The ability to perform this sensitivity analysis for any kind of lens shape is crucial for realistic manufacturing, where there is always error in the control parameters and the code can thus be utilized to define the required manufacturing tolerances.

\begin{figure}
	\centering
        \includegraphics[width=\linewidth]{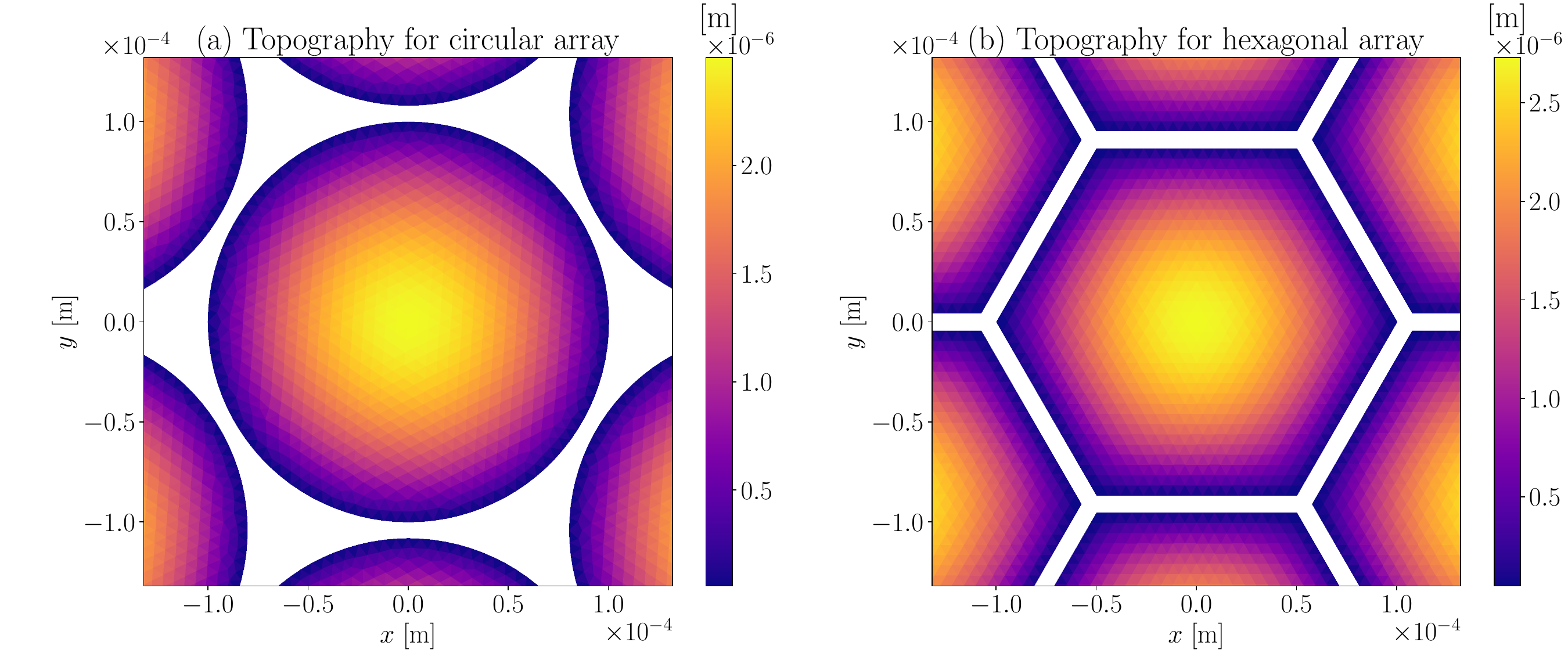}
    \caption{
        Investigation of microlens array geometries. 
        (a) A microlens array based on circular boundaries yield perfectly spherical lenslets, but the geometry inherently limits the fill factor, as indicated by the gaps.
        (b) Hexagonal boundary conditions yield high fill factors at the cost of non-spherical lens topographies.
	Using our solver, we are able to simulate the optical surfaces of such Fluidic Shaping micro-lenses and investigate the trade off between shape and fill factor.
    }
	\label{fig:lenslet}
\end{figure}
Another example for the application of our solver is in the context of micro-lens arrays, which are used in a variety of applications including projection systems, fiber coupling, beam homogenization, and wavefront sensors.
The \emph{fill factor} of such arrays, defined as the ratio of total lenslets area to the area of the entire component, is a principle design consideration and is generally desired to be maximized. 
In order to fabricate such arrays using Fluidic Shaping, one can envision an array of holes into which liquid is deposited.
Such holes can be fabricated using standard micro-fabrication techniques, which generally offer precise control over the footprint with limited control over the boundary height.
Figure \ref{fig:lenslet}(a) presents an array of spherical micro-lenses formed using Fluidic Shaping over such an arrangement of circular micro-holes.
In this design, the circular lenslets are perfectly spherical, but will necessarily leave gaps in the resulting component, limiting the achievable fill factor. 
In contrast, choosing a hexagonal footprint for the boundary conditions, as shown in Figure \ref{fig:lenslet}(b), eliminates the geometrical limitation on the fill factor, and allows the lenslets to tile the planar substrate.
However, the lack of height control on the boundary forces the boundary condition to remain flat, hence preventing the formation of perfectly spherical micro-lenses which in turn reduces the optical quality of the component.
With our code, we can compute the resulting hexagonal micro-lenses and assess their optical properties. 
Specifically, this example also demonstrates our ability to solve the Fluidic Shaping problem over domains with non-smooth boundaries by applying the sharp-point transformation from Appendix \ref{app:dof_trans}. 

\section{Conclusions}
In this work, we presented the mathematical foundation and a numerical solution to the Fluidic Shaping problem over arbitrary domains.
We formulated a general mathematical description of the problem, and established a representative problem, \eqref{eq:repr-prob}, which describes the Fluidic Shaping method for optical applications.
We developed a numerical approach based on reduced quintic finite elements that yields high accuracy solutions for the optical surfaces and their derivatives, and showed how to deform the elements to eliminate geometry mismatch errors while correctly imposing the boundary conditions.
Finally, we verified our solver, compared it to established methods, and demonstrated its capabilities using various examples.

The code developed in this work may serve as the foundation for future studies on the optimal design of  optical components created by Fluidic Shaping.
For example, one may wish to develop an \emph{inverse solver} that determines the boundary conditions, immersion liquid density (i.e., Bond number), and lens volume required to obtain the best approximation of a desired optical surface.
The implementation of such an inverse solver would likely require the repeated solution of the \emph{direct problem} while  modifying the boundary conditions and liquid parameters until an optimal set is found.

Finally, the implementation of deformed reduced quintic finite elements presented in this work could be useful beyond the specific problem studied here.
It could be applied to other mechanical problems, both static and dynamic, whose spatial behavior is described by high-order differential operators.
Prominent examples include the time evolution of thin liquid films and the deformation of thin plates, which are represented by fourth order spatial derivatives.
Obtaining a conforming finite elements formulation for such problems requires $H^2$ continuity, and the reduced quintic elements were designed to represent such highly continuous solutions.
The approach we outline in this work could thus potentially be used to obtain solutions for such problems over arbitrary domains.

\section*{Code availability}
The source code for to develop and verify our solver is available at \url{https://github.com/Fluidic-Technologies-Laboratory/vlm} and licensed under the PolyForm Noncommercial License 1.0.0.

\section*{Acknowldegements}
Funded by the European Union (ERC, Fluidic Shaping, 101044516). Views and opinions expressed are however those of the author(s) only and do not necessarily reflect those of the European Union or the European Research Council Executive Agency. Neither the European Union nor the granting authority can be held responsible for them.

\bibliographystyle{ieeetr}
\bibliography{LensesArticle}

\appendix
\newpage
\section{Parent element formulation for the reduced quintic finite element}\label{app:element_formulation}
\subsection{Constructing the shape functions}
\begin{figure}
	\centering
	\includegraphics[width=0.7\linewidth]{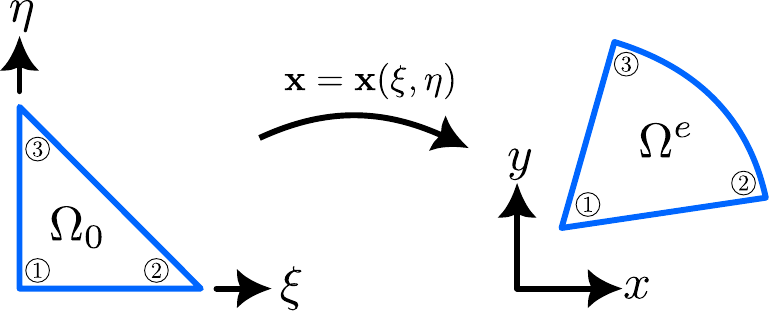}
    \caption{Illustration of the transformation from the parent element to the physical element, for the most generic element considered.}
	\label{fig:element_transformation}
\end{figure}
This supplementary provides the technical details of our implementation of the reduced quintic finite element using the parent element approach.\cite{kirbyGeneralApproachTransforming2018}
The reduced quintic elements are triangular elements with three nodes, $a=1,2,3$ at the vertices. For each node there are  six degrees of freedom,  $r=1,2,3,4,5,6$ , which correspond to the solution value and derivatives at that node, i.e., $u, u_{x}, u_{y}, u_{xx}, u_{yy}, u_{xy}$, respectively.
We denote by $\phi_{(ar)}^e$ the shape function of the reduced quintic element $e$ that corresponds to node $a$ and degree of freedom $r$; meaning that the shape function has unity value at this node and degree of freedom, and it has zero value at all other degrees of freedom.
For example, the $\phi_{(24)}^e$ shape function is associated with node a=2 of the element and its  value and derivatives all vanish at nodes $a=1$ and $3$ . At node $a=2$ they all vanish other than the 4th degree of freedom - the second derivative in the $x$ direction,  $\phi_{(24),xx}^e$, which is one there.
As already established in \eqref{eq:phi_dof_der}, we write $\phi_{(ar),s}^e$ to denote derivation with respect to degree of freedom $s$.
For example, $\phi_{(24),2}^e$ is the $x$ derivative of the shape function.
Using this notation we can express the  shape function relation to a certain degree of freedom using Kroneker's delta, as also described by \eqref{eq:condition_1} in the main text,
\begin{equation}
    \tag{\ref{eq:condition_1}}
    \phi_{(ar),s}(x_b, y_b) = \delta_{(ar)(bs)} = \begin{dcases} 1, &a=b\ \RR{and}\ r=s\\ 0, &\RR{else}\end{dcases}.
\end{equation}

To construct the shape functions that uphold \eqref{eq:condition_1}, we introduce the idea of the \emph{parent element}.
The general idea of the approach is to define a simple element, whose geometry can be mapped onto any physical element in the mesh using an invertible mapping. That way we are able to develop generic computational subroutines, that need only take into account the unique mapping of each element.
Figure \ref{fig:element_transformation} depicts this idea for a deformed triangular element.
The parent element domain, denoted by $\Omega_0$, is a right-isosceles of unit side length, and throughout this section we assume that there exists an invertible mapping, $\vb{x}(\vuxi):\Omega_0\rightarrow\Omega^e$, that maps each point $\vuxi=\qty{\xi,\eta}^T$ in $\Omega_0$ to a point $\vb{x}=\qty{x,y}^T$ in the physical element domain, $\Omega^e$.
Substituting the parent element mapping to the shape function, we can describe the shape function on the parent element as,
\begin{equation}
    \tilde{\phi}_{(ar)}(\vuxi) := \phi_{(ar)}(\vb{x}(\vuxi)).
\end{equation}
To impose condition \pref{eq:condition_1} through the parent element shape function, we first expand $\phi_{(ar),s}$ using the chain rule as follows
\begin{equation}
	\begin{Bmatrix}
        \phi_{(ar)}(\vb{x}) \\ \phi_{(ar),x}(\vb{x}) \\ \phi_{(ar),y}(\vb{x}) \\ \phi_{(ar),xx}(\vb{x}) \\ \phi_{(ar),yy}(\vb{x}) \\ \phi_{(ar),xy}(\vb{x})
	\end{Bmatrix} = 
	\begin{bmatrix}
		1 &0 &0 &0 &0 &0 \\
		0 &\xi_x &\eta_x &0 &0 &0 \\
		0 &\xi_y &\eta_y &0 &0 &0 \\
		0 &\xi_{xx} &\eta_{xx} &\xi_x^2 &\eta_x^2 & 2\xi_x\eta_x\\
		0 &\xi_{yy} &\eta_{yy} &\xi_y^2 &\eta_y^2 & 2\xi_y\eta_y\\
		0 &\xi_{xy} &\eta_{xy} &\xi_x\xi_y &\eta_x\eta_y & \xi_x\eta_y + \xi_y\eta_x
	\end{bmatrix}
	\begin{Bmatrix}
        \tilde{\phi}_{(ar)}(\vuxi) \\ \tilde{\phi}_{(ar),\xi}(\vuxi) \\ \tilde{\phi}_{(ar),\eta}(\vuxi) \\ \tilde{\phi}_{(ar),\xi\xi}(\vuxi) \\ \tilde{\phi}_{(ar),\eta\eta}(\vuxi) \\ \tilde{\phi}_{(ar),\xi\eta}(\vuxi)
	\end{Bmatrix}.
\end{equation}
We denote the components of the chain rule matrix as $\BB_{sS}$ and the components of the vector function which it multiplies as $\tilde{\phi}_{(ar),S}(\vuxi)$.
By plugging in the chain-rule, \eqref{eq:condition_1} becomes
\begin{equation}\label{eq:cond1-xi}
	\sum_{S=1}^{\Ndf} \BB_{sS} \tilde{\phi}_{(ar),S}\eval_{\vuxi_b}
    = \delta_{(ar)(bs)}.
\end{equation}
Note that for any $a$ and $r$, \eqref{eq:cond1-xi} defines $18$ equations, six ($s=1,...,6$) for each node ($b=1,2,3$).

To proceed, we need to make a choice regarding the form of $\tilde{\phi}_{(ar)}$.
As their name suggest, the reduced quintic shape functions are \emph{quintic} polynomials in $x$ and $y$.
In cases where $\vb{x}(\vuxi)$ is an affine transformation, as in Figure \ref{fig:mesh}(b) and (c), then the parent shape functions $\tilde{\phi}_{(ar)}$ are also quintic polynomials, and we can express them as a linear combination of $\xi,\eta$ monomials, i.e.,
\begin{equation}\label{eq:poly-def}
    \tilde{\phi}_{(ar)}(\vuxi) = \bar{\vuxi}^T\vub_{(ar)}
\end{equation}
where 
\begin{multline*}
	\bar{\vuxi}(\vuxi) = \{1, \xi, \eta, \xi^2, \xi\eta, \eta^2, \xi^3, \xi^2\eta , \xi\eta^2, \eta^3, \\ \qquad\qquad \xi^4, \xi^3\eta, \xi^2\eta^2, \xi\eta^3, \eta^4, \xi^5, \xi^4\eta, \xi^3\eta^2, \xi^2\eta^3, \xi\eta^4, \eta^5\}^T,
\end{multline*}
and where $\vub_{(ar)}$ is the vector of $21$ coefficients of the quintic polynomial.
To keep things simple, we use this definition for $\tilde{\phi}_{(ar)}$ even when $\vb{x}(\vuxi)$ is not affine.
However, the corresponding physical shape functions, $\phi_{(ar)}$, may no longer be quintic polynomials.
This may deteriorate the approximation error, but as we saw in the main text, the overall error is still acceptable.
Thus, substituting to \eqref{eq:cond1-xi} the polynomial definition from \eqref{eq:poly-def} we obtain
\begin{equation}\label{eq:cond1-xi-poly}
    \sum_{S=1}^{\Ndf} \BB_{sS} \qty( \bar{\vuxi}_{,S}^{T}\eval_{\vuxi_b}\vub_{(ar)} ) = \delta_{(ar)(bs)}.
\end{equation}

As already established, \eqref{eq:cond1-xi-poly} is a system of $18$ equations for the unknown vector $\vub_{(ar)}$.
However, the vector $\vub_{(ar)}$ has $21$ unknown components, meaning that three more equations are required.
The final $3$ components are determined by setting the normal derivatives of the shape function along each edge to vary as a cubic polynomial.
To impose this condition for some edge, we calculate the normal derivatives at the nodes of that edge, and use them to construct a cubic interpolation for the normal derivative along the edge.
Then, at the midpoint of the edge, we set the normal derivative of the shape function to the interpolated value.
For cases where $\vb{x}(\vuxi)$ is affine, setting this condition makes the normal derivative of the shape function a cubic polynomial along that edge.
Where $\vb{x}(\vuxi)$ is not affine, then $\phi_{(ar)}$ are not necessarily polynomial, but this condition is still sufficient to find the remaining coefficients.

To formulate this condition for any of the edges, we refer to an edge which connects any two nodes $b$ and $c$ in the physical element in Figure \ref{fig:element_transformation}, where the direction $b\rightarrow c$ is agreed to be in the anti-clockwise direction.
Let $\vb{x}_{bc}(t)$ be the parameterization of this edge, with $t\in[0,T]$ as the parameter, and $\vb{x}_b=\vb{x}_{bc}(0)$ and $\vb{x}_c=\vb{x}_{bc}(T)$ as the endpoints.
Note that we do not assume that the parameterization is linear as the edge could be curved.
The normal derivative of a shape function on this edge is
\begin{equation}
    \phi_{(ar),n}(\vb{x}_{bc}(t)) = \phi_{(ar),i}\eval_{\vb{x}_{bc}(t)} \hat{n}_i(t),
\end{equation}
with implicit summation over the index $i$, and the normal direction is given by
\begin{equation}
    \hat{\vb{n}}(t) := \frac{-y_{bc}'(t)\vb{e}_x + x_{bc}'(t)\vb{e}_y}{\sqrt{(x_{bc}')^2 + (y_{bc}')^2}} = \hat{n}_i(t)\vb{e}_i.
\end{equation}
The condition on the normal derivative can thus be stated as
\begin{multline}\label{eq:C2}	
	\phi_{(ar),n}(\vb{x}_{bc}(T/2)) = \frac{1}{2}\qty[\phi_{(ar),n}(\vb{x}_b) + \phi_{(ar),n}(\vb{x}_c)] \\ +\frac{T}{8}\qty[\qty(\dv{t}\phi_{(ar),n})(\vb{x}_b) - \qty(\dv{t}\phi_{(ar),n})(\vb{x}_c)].
\end{multline}
Note that each of the directional derivative terms, i.e. $\phi_{(ar),n}(\vb{x}_{bc}(t))$, in the above expression are calculated at the local normal direction at that point, which is important when the edge is curved.

Next, we use the chain rule to expand the derivative terms, namely,
\begin{equation}\label{eq:phi_n}
	\phi_{(ar),n} = \phi_{(ar),i}\hat{n}_{i} = \phi_{(ar),x}\hat{n}_x + \phi_{(ar),y}\hat{n}_y,
\end{equation}
and
\begin{equation}\label{eq:d_dt_phi_n}
    \dv{t}\qty(\phi_{(ar),n}) = \phi_{(ar),ij}x_{bc,j}'\hat{n}_i + \phi_{(ar),i}\hat{n}_i',
\end{equation}
where an implicit summation is assumed over the indices $i$ and $j$, and where the derivative of the normal vector is
\begin{equation}
    \hat{\vb{n}}' = - \frac{\vb{x}_{bc}'\vdot\vb{x}_{bc}''}{\vb{x}_{bc}'\vdot\vb{x}_{bc}'} \hat{\vb{n}} + \frac{-y_{bc}''\vb{e}_x + x_{bc}''\vb{e}_y}{(\vb{x}_{bc}'\vdot\vb{x}_{bc}')^{1/2}}.
\end{equation}
The final condition on the normal derivative is obtained by substituting \pref{eq:phi_n} and \pref{eq:d_dt_phi_n} into \eqref{eq:C2}, and noting that $\phi_{(ar)}$ and its derivative vanish at most of the evaluation points in accordance with \eqref{eq:condition_1}.
Thus, we obtain the following compact expression
\begin{equation}\label{eq:C2-short}
    \phi_{(ar),n}(\vb{x}_{bc}(T/2)) = \alpha_{abc}\frac{T}{8}\begin{Bmatrix}
			0 \\
			\alpha_{abc}4T^{-1} \hat{n}_x + \hat{n}_x' \\
			\alpha_{abc}4T^{-1} \hat{n}_y + \hat{n}_y' \\
			x'\hat{n}_x \\
			y'\hat{n}_y \\
			x'\hat{n}_y + y'\hat{n}_x
		\end{Bmatrix}\eval_{\vb{x}_a},
\end{equation}
where
\begin{equation}
    \alpha_{abc} = 
    \begin{dcases}
        1, &a=b\\
        -1, &a=c\\
        0, &\RR{else}\\
    \end{dcases}.
\end{equation}
Finally, the left hand side of \eqref{eq:C2} is expanded as 
\begin{equation}\label{eq:phi_xi_mid}
	\phi_{(ar),n}(\vb{x}(T/2)) = \begin{Bmatrix}
		\tilde{\phi}_{(ar),\xi} & \tilde{\phi}_{(ar),\eta}
	\end{Bmatrix}
	\begin{bmatrix}
		\xi_{,x} &\xi_{,y} \\
		\eta_{,x} &\eta_{,y}
	\end{bmatrix}
	\begin{Bmatrix}
		\hat{n}_x \\
		\hat{n}_y
	\end{Bmatrix}\eval_{t=\frac{T}{2}}.
\end{equation}
The system of $18$ equations \pref{eq:cond1-xi} together with $3$ additional equations for each of the edges (\pref{eq:C2-short} with \pref{eq:phi_xi_mid}), uniquely determine the $21$ parameters, namely $\vub_{(ar)}$, that define the shape function $\tilde{\phi}_{(ar)}$.
This process is then repeated for each of the $18$ shape functions of the element, and the result is cached for the computation.

\subsection{Parent Element Transformation}
All the formulas in the previous section relied on the existence of the parent transformation $\vb{x}(\vuxi):\Omega_0\rightarrow\Omega^e$, and used properties of its inverse, such as $\xi_{,x}$ and $\eta_{,y}$, for the shape functions calculations.
Here we present the transformations used in our solver, and give the necessary inversion formulas.

\paragraph{Affine transformation.}
Whenever we consider an element $\Omega^e$ which is only stretched and rotated with respect to the parent $\Omega_0$ (for example consider Figure \ref{fig:mesh}(b) and (c)), then $\vb{x}(\vuxi)$ is the simple following affine transformation
\begin{equation}
	\begin{Bmatrix}
		x(\xi,\eta)\\ y(\xi,\eta)
	\end{Bmatrix} = 
	\begin{bmatrix}
		x_2-x_1 &x_3-x_1\\
		y_2-y_1 &y_3-y_1
	\end{bmatrix}\begin{Bmatrix}
		\xi \\ \eta
	\end{Bmatrix} + 
	\begin{Bmatrix}
		x_1 \\ y_1
	\end{Bmatrix}.
\end{equation}

\paragraph{Nonlinear transformation.}
Whenever the shape of the element is a deformed triangle, such as in Figure \ref{fig:mesh}(d), we use the following transfinite transformation inspired by Barnhill et al.\cite{BarnhillBirkhoffGordon1973},
\begin{equation}\label{eq:simple-transfinite}
	\vb{x}(\vuxi) = \frac{1}{2}\qty\Bigg[\vb{U}_1(\vuxi) + \vb{U}_2(\vuxi) + \vb{U}_3(\vuxi) - \qty\Big(\vb{x}_1 + \qty(\vb{x}_2-\vb{x}_1)\xi + \qty(\vb{x}_3-\vb{x}_1)\eta)],
\end{equation}
where
\begin{align}
	\vb{U}_1(\vuxi) &= \qty(1 - \xi - \eta)\vb{F}_{31}(1-\eta) + \xi\vb{F}_{23}(\eta), \\
	\vb{U}_2(\vuxi) &= \xi\vb{F}_{12}(\xi+\eta) + \eta\vb{F}_{31}(1-\xi-\eta), \\
	\vb{U}_3(\vuxi) &= \qty(1-\xi-\eta)\vb{F}_{12}(\xi) + \eta \vb{F}_{23}(1-\xi),
\end{align}
and where $\vb{F}_{ab}(t)$ is a parameterization with $t\in[0,1]$ of the element edge in the physical domain which connects node $a$ to node $b$ (in that orientation).
This kind of transformation maps specific lines in $\vuxi$ space to specific curves in $\vb{x}$, while continuously blending the rest of the space.\cite{gordonConstructionCurvilinearCoordinate1973,gordonTransfiniteMappingsTheir1982}
In our case, the transformation maps the straight edges of the right isosceles $\Omega_0$ to the potentially curved edges of the element $\Omega^e$, where the $\vb{F}_{ab}(t)$ are curve parameterizations.
Note, that if all of $\vb{F}_{ab}(t)$ are linear in $t$ then the transfinite mapping is reduced to the affine mapping.
However, in terms of computational performance, it is worthwhile to make the distinction.

\paragraph{Transformation inversion.}
As we just saw, the parent element transformation is formulated from $\vuxi$ to $\vb{x}$; however, for the shape function calculations we use the derivatives of the inverse mapping.
For example, to find the values of  $\xi_{,x}$ and $\eta_{,x}$, one needs to solve the following system of equations,
\begin{align}
	& x_{,x} = 1 = x_{,\xi}\xi_{,x} + x_{,\eta}\eta_{,x}, \\
	& y_{,x} = 0 = y_{,\xi}\xi_{,x} + y_{,\eta}\eta_{,x}.
\end{align}
A similar system is obtained for $\xi_{,y}$ and $\eta_{,y}$.
Overall we get the following systems of equations,
\begin{equation}
	\begin{bmatrix}
		x_{,\xi} &x_{,\eta} \\
		y_{,\xi} &y_{,\eta}
	\end{bmatrix}
	\begin{bmatrix}
		\xi_{,x} & \xi_{,y} \\
		\eta_{,x} &\eta_{,y}
	\end{bmatrix} = 
	\begin{bmatrix}
		1 & 0 \\
		0 & 1
	\end{bmatrix}.
\end{equation}
After obtaining the first derivatives of the inverse transform, we can find the second derivatives.
For example, for $\xi_{,xx}$ and $\eta_{,xx}$ one needs to solve
\begin{align}
	x_{,xx} = 0 = x_{,\xi\xi}\xi_{,x}^2 + x_{,\eta\eta}\eta_{,x}^2 + 2 x_{,\xi\eta}\xi_{,x}\eta_{,x} + x_{,\xi}\xi_{,xx} + x_{,\eta}\eta_{,xx}, \\
	y_{,xx} = 0 = y_{,\xi\xi}\xi_{,x}^2 + y_{,\eta\eta}\eta_{,x}^2 + 2 y_{,\xi\eta}\xi_{,x}\eta_{,x} + y_{,\xi}\xi_{,xx} + y_{,\eta}\eta_{,xx},
\end{align}
and similar equations can be obtained for $\xi_{,yy},\eta_{,yy}$ and $\xi_{,xy},\eta_{,xy}$.
In matrix form, these gives
\begin{multline}
	\begin{bmatrix}
		x_{,\xi} &x_{,\eta} \\
		y_{,\xi} &y_{,\eta}
	\end{bmatrix}
	\begin{bmatrix}
		\xi_{,xx} &\xi_{,yy} &\xi_{,xy} \\ \eta_{,xx} &\eta_{,yy} &\eta_{,xy}
	\end{bmatrix} \\ = -
	\begin{bmatrix}
		x_{,\xi\xi} & x_{,\eta\eta} & x_{,\xi\eta} \\
		y_{,\xi\xi} & y_{,\eta\eta} & y_{,\xi\eta}
	\end{bmatrix}
	\begin{bmatrix}
		\xi_{,x}^2 & \xi_{,y}^2 & \xi_{,x}\xi_{,y} \\ \\
		\eta_{,x}^2 & \eta_{,y}^2 & \eta_{,x}\eta_{,y} \\ \\ 2\xi_{,x}\eta_{,x} & 2\xi_{,y}\eta_{,y} & (\xi_{,x}\eta_{,y} + \xi_{,y}\eta_{,x})
	\end{bmatrix}.
\end{multline}

\newpage
\section{Imposing Dirichlet Boundary conditions}\label{app:dof_trans}
\begin{figure}
    \centering
    \includegraphics[width=0.9\linewidth]{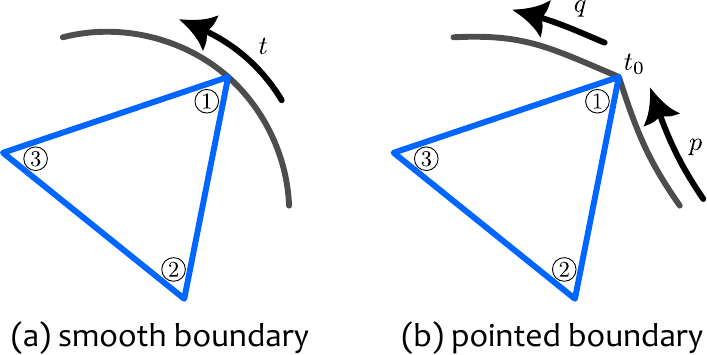}
    \caption{Illustration of the two possible boundary conditions types: (a) smooth boundary and (b) pointed boundary.}
    \label{fig:boundary_conditions}
\end{figure}
In this supplementary we show how to impose the pinning (Dirichlet) boundary conditions on nodes which are either on a smooth part of the boundary (see Figure \ref{fig:boundary_conditions}(a)), or on a sharp point of the boundary (see Figure \ref{fig:boundary_conditions}(b)). 

\subsection{Boundary conditions on a smooth boundary}
Let $\bar{x}=\bar{x}(t)$, $\bar{y}=\bar{y}(t)$ and $\bar{u}=\bar{u}(t)$ represent a smooth parameterization of the boundary's curve in three dimensions, and denote by $\bar{x}_{t}$ differentiation with respect to $t$.
For a smooth curve, we assume that the tangent is well-defined, i.e. that $\bar{x}_t^2+\bar{y}_t^2 > 0$ at any point on the curve.
Therefore, to impose the pinning boundary condition we require that
\begin{equation}
	\begin{split}
		&u = \bar{u},\\
		&\dv{t}\qty(u(\bar{x}(t),\bar{y}(t))) = \bar{u}_{t},\\
		&\dv[2]{t}\qty(u(\bar{x}(t),\bar{y}(t))) = \bar{u}_{tt}.
	\end{split}
\end{equation}
By the chain rule, this becomes
\begin{equation}
	u_{x}\bar{x}_{t} + u_{y}\bar{y}_{t} = \bar{u}_{t},
\end{equation}
and
\begin{equation}
	u_{x}\bar{x}_{tt} + u_{y}\bar{y}_{tt} + u_{xx}\bar{x}_{t}^2 + u_{yy}\bar{y}_{t}^2 + 2u_{xy}\bar{x}_{t}\bar{y}_{t} = \bar{u}_{tt}.
\end{equation}
We now replace $u$ with $u^h$ and rearrange this expression into a matrix form using the nodal degrees of freedom notation
\begin{equation}\label{eq:BC-transform-1}
	\begin{bmatrix}
		1 &0 &0 &0 &0 &0\\
		0 &\bar{x}_{t} &\bar{y}_{t} &0 &0 &0\\
		0 &\bar{x}_{tt} &\bar{y}_{tt} &\bar{x}_{t}^2 &\bar{y}_{t}^2 &2\bar{x}_{t}\bar{y}_{t}
	\end{bmatrix}
	\begin{Bmatrix}
        d_{(a1)}^e \\ d_{(a2)}^e \\ d_{(a3)}^e \\ d_{(a4)}^e \\ d_{(a5)}^e \\ d_{(a6)}^e
	\end{Bmatrix} = 
	\begin{Bmatrix}
		\bar{u}\\ \bar{u}_{t}\\ \bar{u}_{tt}
	\end{Bmatrix}.
\end{equation}
Beyond these three equations, there are no other requirements that could be used to fix the other nodal degrees of freedom.
Wu et al. \cite{wuC1ConformingQuadrilateral2020}, thus suggested to view \eqref{eq:BC-transform-1} as part of a transformation from $\vb{d}_a^e$ to $\vb{d}_a^{e*}$, where the first degrees of freedom are fixed as $d_{(a1)}^{e*}=\bar{u}$, $d_{(a2)}^{e*}=\bar{u}_{t}$, $d_{(a3)}^{e*}=\bar{u}_{tt}$ and  the rest of the nodal degrees of freedom ($d_{(a4)}^{e*}$, $d_{(a5)}^{e*}$ and $d_{(a6)}^{e*}$) remain variable.
The remaining three transformation equations are thus somewhat arbitrary, but should be chosen to avoid singularity.
We find that the following transformation satisfies well these requirements, 
\begin{equation}\label{eq:BC-transform-2}
	\begin{bmatrix}
		1 &0 &0 &0 &0 &0\\
		0 &\bar{x}_{t} &\bar{y}_{t} &0 &0 &0\\
		0 &\bar{x}_{tt} &\bar{y}_{tt} &\bar{x}_{t}^2 &\bar{y}_{t}^2 &2\bar{x}_{t}\bar{y}_{t}\\
		0 &-\bar{y}_{t} &\bar{x}_{t} &0 &0 &0\\
		0 &0 &0 &\bar{y}_{t}^2 &-\bar{x}^2_{t} &0\\
		0 &0 &0 &0 &0 &1
	\end{bmatrix}
	\begin{Bmatrix}
        d_{(a1)}^e \\ d_{(a2)}^e \\ d_{(a3)}^e \\ d_{(a4)}^e \\ d_{(a5)}^e \\ d_{(a6)}^e
	\end{Bmatrix} = 
	\begin{Bmatrix}
        d_{(a1)}^{e*} \\ d_{(a2)}^{e*} \\ d_{(a3)}^{e*} \\ d_{(a4)}^{e*} \\ d_{(a5)}^{e*} \\ d_{(a6)}^{e*}
	\end{Bmatrix}.
\end{equation}
We denote the transformation matrix by $\vb{T}^{s}$ where $s$ is for \emph{smooth}.
As a final note, the determinant is
\begin{equation}
	\det(\vb{T}^s) = \bar{x}_{t}^6 + \bar{x}_{t}^4\bar{y}_{t}^2 + \bar{x}_{t}^2\bar{y}_{t}^4 + \bar{y}_{t}^6,
\end{equation}
hence $\vb{T}^s$ is invertible as long as $\bar{x}_t$ and $\bar{y}_{t}$ aren't both zero, which is satisfied by the above-mentioned requirement on the parameterization.
Note that in many calculations the inverse transformation appears, and we denote it by $\vb{Q}^{s} = (\vb{T}^{s})^{-1}$.

\subsection{Boundary conditions at a sharp point}
In cases where there are sharp points along the boundary, the transformation from \eqref{eq:BC-transform-2} doesn't apply, due to the discontinuity of the derivatives of the boundary parameterization at the sharp point.
Suppose that $t=t_0$ is a sharp point in the parameterization.
For simplicity of notation, let $p:=t$ for $t<t_0$ and let $q:=t$ for $t>t_0$.
At $t=t_0$, from the pinning boundary condition, we have the data on $\bar{u}$, $\bar{u}_{,p}$, $\bar{u}_{,pp}$, $\bar{u}_{,q}$ and $\bar{u}_{,qq}$. Thus, by the chain rule, we get
\begin{equation}\label{eq:pointed_bc_1st}
	\begin{split}
		\bar{u}_{,p} &= u_{,x} \bar{x}_{,p} + u_{,y} \bar{y}_{,p},\\
		\bar{u}_{,q} &= u_{,x} \bar{x}_{,q} + u_{,y} \bar{y}_{,q},
	\end{split}
\end{equation}
and
\begin{equation}\label{eq:pointed_bc_2st}
	\begin{split}
		\bar{u}_{,pp} &= u_{,x}\bar{x}_{,pp} + u_{,y}\bar{y}_{,pp} + u_{,xx}\bar{x}_{,p}^2 + u_{,yy}\bar{y}_{,p}^2 + 2u_{,xy}\bar{x}_{,p}\bar{y}_{,p},\\
		\bar{u}_{,qq} &= u_{,x}\bar{x}_{,qq} + u_{,y}\bar{y}_{,qq} + u_{,xx}\bar{x}_{,q}^2 + u_{,yy}\bar{y}_{,q}^2 + 2u_{,xy}\bar{x}_{,q}\bar{y}_{,q},\\
	\end{split}
\end{equation}
Equations \pref{eq:pointed_bc_1st} and \pref{eq:pointed_bc_2st}, together with $u=\bar{u}$, provide five out of six required conditions on the nodal degrees of freedom.
As in the smooth case, we seek to define a transformation from $\vb{d}_a^e$ to $\vb{d}_a^{e*}$.
In this case we only require one more transformation equation.
We here choose to use the mixed derivative relation 
\begin{equation}
	\bar{u}_{,pq} = u_{,x}\bar{x}_{,pq} + u_{,y}\bar{y}_{,pq} + u_{,xx}\bar{x}_{,p}\bar{x}_{,q} + u_{,yy}\bar{y}_{,p}\bar{y}_{,q} + (\bar{x}_{,p}\bar{y}_{,q} + \bar{x}_{,q}\bar{y}_{,p})u_{,xy}.
\end{equation}
Overall, the transformation for ``sharp" points on the boundary is
\begin{equation}\label{eq:BC-sharp}
	\begin{bmatrix}
		1 &0 &0 &0 &0 &0\\
		0 &\bar{x}_{,p} &\bar{y}_{,p} &0 &0 &0\\
		0 &\bar{x}_{,q} &\bar{y}_{,q} &0 &0 &0\\
		0 &\bar{x}_{,pp} &\bar{y}_{,pp} &\bar{x}_{,p}^2 &\bar{y}_{,p}^2 &2\bar{x}_{,p}\bar{y}_{,p}\\
		0 &\bar{x}_{,qq} &\bar{y}_{,qq} &\bar{x}_{,q}^2 &\bar{y}_{,q}^2 &2\bar{x}_{,q}\bar{y}_{,q}\\
		0 &\bar{x}_{,pq} &\bar{y}_{,pq} &\bar{x}_{,p}\bar{x}_{,q} &\bar{y}_{,p}\bar{y}_{,q} &\bar{x}_{,p}\bar{y}_{,q} + \bar{x}_{,q}\bar{y}_{,p}\\
	\end{bmatrix}
	\begin{Bmatrix}
        d_{(a1)}^e \\ d_{(a2)}^e \\ d_{(a3)}^e \\ d_{(a4)}^e \\ d_{(a5)}^e \\ d_{(a6)}^e
	\end{Bmatrix} = 
	\begin{Bmatrix}
        d_{(a1)}^{e*} \\ d_{(a2)}^{e*} \\ d_{(a3)}^{e*} \\ d_{(a4)}^{e*} \\ d_{(a5)}^{e*} \\ d_{(a6)}^{e*}
	\end{Bmatrix}.
\end{equation}
We denote the transformation matrix by $\vb{T}^{p}$ where $p$ is for the \emph{sharp point}.
Finally, we note that the determinant is
\begin{equation}\label{eq:det_Tp}
	\det(\vb{T}^p) = (\bar{x}_{,p}\bar{y}_{,q}-\bar{x}_{,q}\bar{y}_{,p})^4,
\end{equation}
hence $\vb{T}^p$ is invertible as long as there is indeed a sharp point at $t=t_0$.
Note that in many calculations the inverse transformation appears, and we denote it by $\vb{Q}^{p} = (\vb{T}^{p})^{-1}$.

\newpage
\section{The Galerkin-Ritz finite element formulation for the Fluidic Shaping problem}\label{app:galerkin_ritz_fe_formulation}
Recall the functional from \eqref{eq:repr-prob} for which we seek a stationary value,
\begin{equation}
    \Pi[u;P] = \int_{\Omega}\qty[\sqrt{1 + u_{,x}^2 + u_{,y}^2} - \frac{1}{2}\Bo u^2] \dd{\Omega} - P \qty(V - \int_{\Omega} u \dd{\Omega}).
    \tag{\ref{eq:repr-prob}}
\end{equation}
Replacing $\Omega$ with $\Omega^h$, and substituting $u^h\in S^h$ in place of $u\in S$ we obtain
\begin{equation}\label{eq:Pi_h}
    \Pi[u^h;P] = \int_{\Omega^h}\qty[\sqrt{1 + (u_{,x}^h)^2 + (u_{,y}^h)^2} - \frac{1}{2}\Bo (u^h)^2] \dd{\Omega} - P \qty(V - \int_{\Omega^h} u^h \dd{\Omega}),
\end{equation}
where we have omitted the numerical error, $E(u-u^h)$, and the geometry mismatch error, $E(\Omega\Delta\Omega^h)$.
Because the approximation $u^h$ is expressed using a finite number of nodal degrees of freedom, then $\Pi[u^h;P]$, which is a \emph{functional} of the unknown function $u^h$, can be written as a \emph{function} of the unknown nodal degrees of freedom $\vb{d}^*$, hereby denoted as $\Pi^h(\vb{d}^{*}, P) := \Pi[u^h,P]$.

To find the extremum of the approximate potential function, $\Pi^h(\vb{d}^*;P)$, we differentiate it with respect to the unknowns and equate to zero, i.e.,
\begin{align}
    &\pdv{\Pi^h(\vb{d}^*; P)}{\vb{d}^*} =\vb{G}^{u} (\vb{d}^*; P) - \vb{F}^u =  \vb{0},\\
    &\pdv{\Pi^h(\vb{d}^*; P)}{P}={G}^{P} (\vb{d}^*; P) -    {F}^P  = 0.
\end{align}
Here, $\vb{G}^u$ and $\vb{G}^P$ refer only to the derivative terms which explicitly depends on the unknowns, $\vb{d}^*$ and $P$, whereas $\vb{F}^u$ and $F^P$ are the constant terms.
As such, for our problem, $\vb{F}^u=\vb{0}$ and  $F^P = V$.

Thus, the goal of our finite elements code is to solve the following nonlinear system of equations
\begin{equation}\label{eq:nonlinear_system}
	\begin{dcases}
		\vb{G}^u(\vb{d}^{*};P) = \vb{F}^u\\
		G^P(\vb{d}^{*}) = F^P
	\end{dcases}.
\end{equation}
Most algorithms for solving nonlinear equations, such as Newton iterations, require the computation of the tangent stiffness matrix $\vb{K}$ (the Jacobian of $\vb{G}$) and the residual vector $\vb{R}=\vb{G}-\vb{F}$, which in our case take the form
\begin{equation}
    \vb{K} = 
    \begin{bmatrix}
		\vb{K}^{uu} &\vb{K}^{up}\\
		\vb{K}^{pu} & 0
	\end{bmatrix} 
    \qq{and}
	\vb{R}^{} = \begin{Bmatrix}
		\vb{R}^{u}\\
		R^{P}
	\end{Bmatrix}.
\end{equation}
Each of these are calculated at the element level and then assembled using the \emph{finite elements assembly process}, denoted in the following by \emph{big A},
\begin{equation}
	\begin{split}
		&\vb{K}^{uu} = \Ass_{e=1}^{N_{el}} \vb{k}^{euu},\quad \vb{K}^{up} = \Ass_{e=1}^{N_{el}} \vb{k}^{eup} ,\quad \vb{K}^{pu} = \Ass_{e=1}^{N_{el}} \vb{k}^{epu} ,\quad \\
		&\vb{R}^{u} = \Ass_{e=1}^{N_{el}} \qty(\vb{f}^{eu}-\vb{g}^{eu}),\quad
		R^P = F^{P} - \sum_{e=1}^{N_{el}} g^{eP}.
    \end{split}
\end{equation}
where $F^P=V$, as previously established.
Finally, the components of the $\vb{k}^{euu}$, $\vb{k}^{epu}$, $\vb{k}^{eup}$, $\vb{g}^{eu}$ and $\vb{f}^{eu}$ arrays are calculated using the following formulas
\begin{equation}\label{eq:FE-form}
	\begin{split}
		&k_{(ar)(bs)}^{euu,\RR{lin.}} = -\int_{\Omega^e}\phi_{(ar)}^{*}\Bo\phi_{(bs)}^{*}\dd{\Omega} + \int_{\Omega^e}\phi_{(ar),j}^{*}\kappa\phi_{(bs),j}^{*}\dd{\Omega} ,\\
		&k_{(ar)(bs)}^{euu} = k_{(ar)(bs)}^{euu,\RR{lin.}} - \int_{\Omega^e} \phi_{(ar),j}^{*} u_{,j}^h\kappa^{\prime} u^h_{,k}\phi_{(bs),k}^{*}\dd{\Omega} ,\\
		&k_{(ar)}^{eup} = \int_{\Omega^e} \phi_{(ar)}^{*}\dd{\Omega}\ ,\quad
		k_{(bs)}^{epu} = \int_{\Omega^e} \phi_{(bs)}^{*}\dd{\Omega},\\
		&g_{(ar)}^{eu} = \sum_{b=1}^{N_{en}} \sum_{s=1}^{\Ndf} k_{(ar)(bs)}^{euu,\RR{lin.}} d_{(bs)}^{*e} + k_{(ar)}^{eup}P,\\
		&g^{eP} = \sum_{b=1}^{N_{en}} \sum_{s=1}^{\Ndf} k_{(bs)}^{epu}d_{(bs)}^{*e},\\ 
		&f_{(ar)}^{eu} = 0,
	\end{split}
\end{equation}
where
\begin{equation}
    \kappa = \qty[1+u_{,i}^hu_{,i}^h]^{-1/2},\qquad 
    \kappa' = \qty[1+u_{,i}^hu_{,i}^h]^{-3/2} = \kappa^3.
\end{equation}

\subsection{Calculating the Element's Arrays}
\paragraph{The $ uu $ contribution to the element's stiffness matrix.}
We start by considering the ``linear" part of the element stiffness matrix.
By transforming the integrals of $\vb{k}^{euu,\RR{lin}.}$ onto the parent element we obtain
\begin{equation}\label{eq:keuu,lin}
	\begin{split}
		k_{(ar)(bs)}^{euu,\RR{lin}.} &= -\int_{\Omega^e} \phi_{(ar)}^{*}\Bo\phi_{(bs)}^{*} \dd{\Omega} + \int_{\Omega^e} \phi_{(ar),j}^{*}\kappa\phi_{(bs),j}^{*}\dd{\Omega}\\
		&= - \underbracket{\int_{\Omega_0} \tilde{\phi}_{(ar)}^{*} \Bo\tilde{\phi}_{(bs)}^{*} j\dd{\xi}\dd{\eta}}_{\RR{(A)}} + \underbracket{\int_{\Omega_0}  \frac{\tilde{\phi}_{(ar),k}^{*} \tilde{\phi}_{(bs),k}^{*}}{\sqrt{1+u^h_{,i}u^h_{,i}}}  j\dd{\xi}\dd{\eta}}_{\RR{(B)}},
	\end{split}
\end{equation}
where $j = \det(\pdv*{\vb{x}}{\vuxi})$ is the Jacobian of the parent element transformation.
The term (A) can be rewritten as
\begin{equation}\label{eq:(A)}
	\begin{split}
		\RR{(A)} &= \int_{\Omega_0} \tilde{\phi}_{(ar)}^{*} \Bo\tilde{\phi}_{(bs)}^{*} j\dd{\xi}\dd{\eta} \\
		&= \int_{\Omega_0} \tilde{\phi}_{(ct)}Q^e_{(ct)(ar)} \Bo\tilde{\phi}_{(dq)}Q^e_{(dq)(bs)} j\dd{\xi}\dd{\eta} \\
		&= \Bo Q_{(ct)(ar)}^e \int_{\Omega_0} \tilde{\phi}_{(ct)} \tilde{\phi}_{(dq)} j\dd{\xi}\dd{\eta} Q_{(dq)(bs)}^e \\
		&= \Bo Q_{(ct)(ar)}^e \beta_{(ct)\mu} \int_{\Omega_0} \bar{\xi}_{\mu}\bar{\xi}_{\nu} j\dd{\xi}\dd{\eta} \beta_{(dq)\nu} Q_{(dq)(bs)}^e,
	\end{split}
\end{equation}
where $\vub_{(ar)}$ and $\bar{\vuxi}$ were given in Appendix \ref{app:element_formulation}, $\vb{Q}^e$ is a $18\times18$ matrix with three $6\times6$ blocks on its main diagonal corresponding to the nodal transformation (given in Appendix \ref{app:dof_trans}) of each node.
Moreover, the summation over $(ct)$, $(dq)$, $\mu$ and $\nu$, are carried out implicitly here and also in everything that follows.

As already established, in the most generic case, the transformation from $\Omega_0$ to $\Omega^e$ need not be linear.
In such case, the integral in \eqref{eq:(A)} is calculated via quadrature (see Appendix \ref{app:GLQ}).
However, as most of the elements in the mesh are internal, their parent transformation is affine, meaning that their Jacobian is constant, and $j$ can be taken out of the integral in \eqref{eq:(A)}.
In such a case, we may calculate it a-priori once.

We now proceed with term (B) in \eqref{eq:keuu,lin}.  We can explicitly expand the numerator using the chain rule,
\begin{equation}
	\begin{split}
		k_{(ar)(bs)}^* := \tilde{\phi}_{(ar),k}^{*} \tilde{\phi}_{(bs),k}^{*} &= Q_{(ct)(ar)}^e \tilde{\phi}_{(ct),k}\tilde{\phi}_{(dq),k} Q_{(dq)(bs)}^e \\
		&= Q_{(ct)(ar)}^e \tilde{\phi}_{(ct),K}\xi_{K,k} \tilde{\phi}_{(dq),L}\xi_{L,k} Q_{(dq)(bs)}^e \\
		&= Q_{(ct)(ar)}^e \beta_{(ct)\mu}\bar{\xi}_{\mu,K} \xi_{K,k} \beta_{(dq)\nu}\bar{\xi}_{\nu,L}\xi_{L,k} Q_{(dq)(bs)}^e.
	\end{split}
\end{equation}
Here $\xi_K$ represents the vector components of $\vuxi=\qty{\xi,\eta}^T$.
We denote by $\square_{,k}$ the differentiation with respect to a physical coordinate $x$ or $y$, and by $\square_{,K}$ the differentiation with respect to a reference coordinate $\xi$ or $\eta$.
In the denominator of term (B) we have $u^h_{,i}u^h_{,i}$ which we calculate as follows
\begin{equation}
	u^h_{,i}u^h_{,i} = \qty(d_{(ct)}^{*e}\tilde{\phi}_{(ct),i}^{*}) \qty(d_{(dq)}^{*e}\tilde{\phi}_{(dq),i}^{*}) = d^{*e}_{(ct)} \qty( \tilde{\phi}_{(ct),i}^{*}\tilde{\phi}_{(dq),i}^{*}) d^{*e}_{(dq)} = d^{*e}_{(ct)}k_{(ct)(dq)}^{*} d^{*e}_{(dq)}.
\end{equation}
Thus, for the (B) term, we get
\begin{equation}
	\RR{(B)} = \int_{\Omega_0} \frac{k^*_{(ar)(bs)}}{\sqrt{1 + d^{*e}_{(ct)}k^*_{(ct)(dq)} d^{*e}_{(dq)}}} j \dd{\xi}\dd{\eta}.
\end{equation}

Next, the nonlinear part of the $ euu $ stiffness matrix is given by
\begin{equation}
	\begin{split}
		&\!\!\!\!\!\!\!\!\!\!\!\!\!\!\!\!\!\!\!\!\!\!\!\!\!\!\!
		\int_{\Omega^e} \phi_{(ar),l}^{*} u^h_{,l}\kappa'u^h_{,k}\phi_{(bs),k}^{*} \dd{\Omega}\\
		&=\int_{\Omega_0} \frac{ \tilde{\phi}_{(ar),l}^{*}\tilde{\phi}_{(ct),l}^{*} d_{(ct)}^{*e}\tilde{\phi}_{(bs),k}^{*}\tilde{\phi}_{(du),k}^{*}d_{(du)}^{*e}}{\qty[1 + u^h_{,i}u^h_{,i}]^{3/2}} j\dd{\xi}\dd{\eta} \\
		&=\int_{\Omega_0} k^*_{(ar)(ct)} d_{(ct)}^{*e} k_{(bs)(dq)}^{*} d_{(dq)}^{*e} \kappa^3 j\dd{\xi}\dd{\eta}.
	\end{split}
\end{equation}
Here, even if $j$ is constant, the integrand is still non-polynomial and we have to use a quadrature to evaluate the integral numerically.

\paragraph{The $ up $ and $ pu $ contribution to the element's stiffness matrix.}
First note that $ \vb{k}^{eup} = \qty(\vb{k}^{epu})^T $. To calculate these matrices recall the fact that $ \sum_{b=1}^{N_{en}}\phi_{(b1)}\equiv1 $ so we get that,
\begin{multline}
	\vb{k}^{eup} = \qty(\vb{k}^{epu})^T = \qty[\int_{\Omega^{e}}\phi_{(ar)}^*\dd{\Omega}] = \qty[\int_{\Omega_0}\tilde{\phi}_{(ar)}^*j\dd{\xi}\dd{\eta}] \\ = \qty[\sum_{b=1}^{N_{en}} \int_{\Omega_0}\tilde{\phi}_{(ar)}^*\tilde{\phi}_{(b1)}^*j\dd{\xi}\dd{\eta}] = :\vb{m}^{\gen}\mathds{1}
\end{multline}
where
\begin{equation}
    m_{(ar)(bs)}^{\gen} := \int_{\Omega_0} \tilde{\phi}_{(ar)}^{*} \tilde{\phi}_{(bs)}^{*} j \dd{\xi}\dd{\eta},
\end{equation}
and where $ \mathds{1} $ is a vector of length $ N_{en}\times\Ndf $ whose components are unity for the degrees-of-freedom $s=1$ and zero otherwise, i.e., it's the values vector for the finite element interpolation of a unit function.

\section{Gauss-Legendre quadrature for triangles}\label{app:GLQ}
Following Cowper \cite{cowperGaussianQuadratureFormulas1973}, to integrate over a domain of a right-triangle with perpendiculars of length one joined at the $(0,0)$ origin, we use the following formula:
\begin{equation}
	\int_{\Omega_0} f(\xi,\eta)\dd{\xi}\dd{\eta} = \frac{1}{2}\sum_{i=1}^{12} w_i f(\xi^*_i, \eta^*_i)
\end{equation}
where
\begin{equation*}
	\begin{array}{c|c|c|c}
        i &w_i & \xi^*_i & \eta^*_i \\\hline\hline
		1  &0.050844906370207  &0.873821971016996 &0.063089014491502 \\
		2  &0.050844906370207  &0.063089014491502 &0.873821971016996 \\
		3  &0.050844906370207  &0.063089014491502 &0.063089014491502 \\
		4  &0.116786275726379  &0.501426509658179 &0.249286745170910 \\
		5  &0.116786275726379  &0.249286745170911 &0.501426509658179 \\
		6  &0.116786275726379  &0.249286745170910 &0.249286745170911 \\
		7  &0.082851075618374  &0.636502499121399 &0.310352451033785 \\
		8  &0.082851075618374  &0.636502499121399 &0.053145049844816 \\
		9  &0.082851075618374  &0.310352451033785 &0.053145049844816 \\
		10 &0.082851075618374  &0.053145049844816 &0.310352451033785 \\
	    11 &0.082851075618374  &0.053145049844816 &0.636502499121399 \\
	    12 &0.082851075618374  &0.310352451033785 &0.636502499121399 \\ \hline
	\end{array}
\end{equation*}

\section{Fluidic Shaping of freeform optical component with small boundary variations}\label{app:small_boundary_variations}
\begin{figure}
	\centering
	\includegraphics[width=\linewidth]{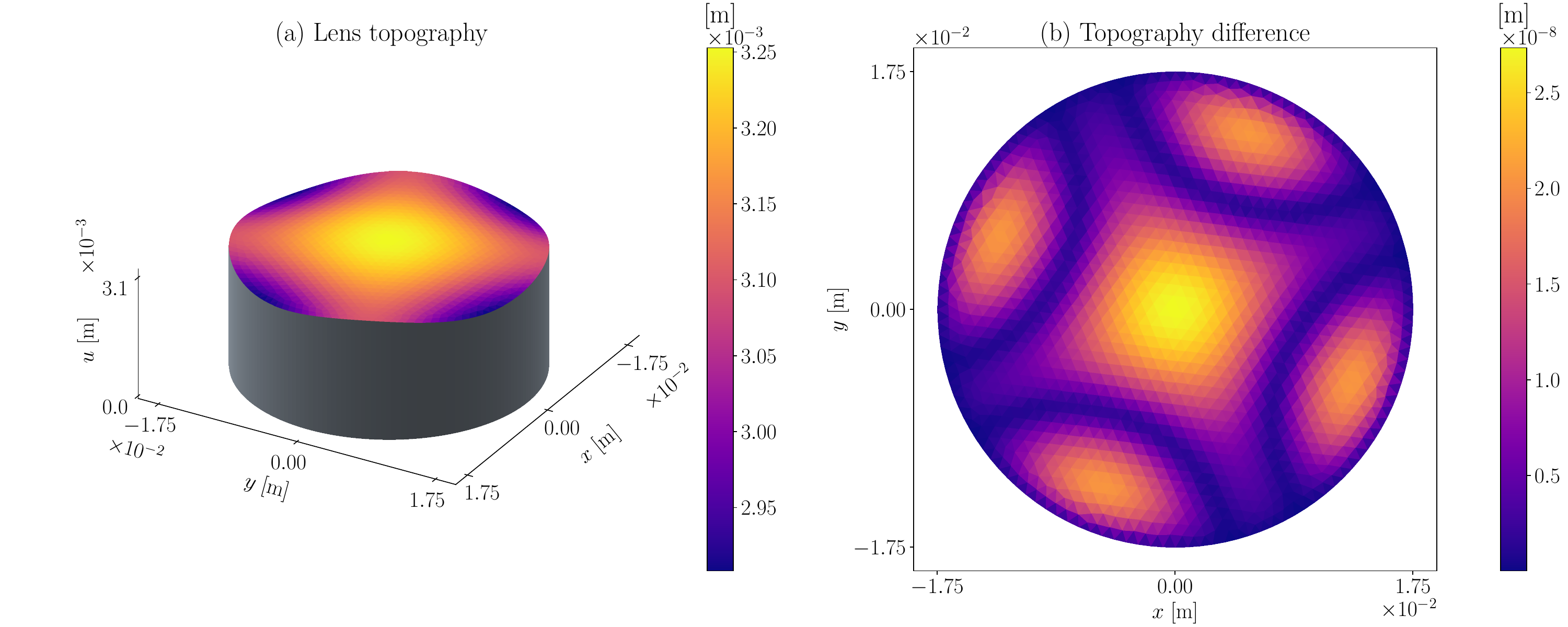}
    \caption{
        The analytical solution of Elgarisi et al.\cite{ElgarisiOptica} can be used to design precision freeform optical components for surfaces with small deformations.
        (a) An orthogonal view of the lens.
        (b) The absolute difference in the topography of the analytic and numeric surfaces.
        This lens was obtained for radius $1.75$ cm, volume $3$ ml, Bond $3$, and height $3 + 0.1\sin(4\theta)$ mm, for $\theta\in[0,2\pi)$.
    }
	\label{fig:elgarisi_100mu}
\end{figure}

In Figure \ref{fig:elgarisi} of the main text we presented a circular freeform optical component with a diameter of $3.5$ cm and a $550$ $\RR{\mu m}$ sinusoidal variation, and compared the numerical solution to the analytical solution of Elgarisi et al.\cite{ElgarisiOptica}
In that example, the magnitude of the sinusoidal variation was too large to satisfy the assumptions of the linear model and we concluded that the analytical result is not sufficiently accurate for precision optics.
In Figure \ref{fig:elgarisi_100mu} we show the same result for a component with a sinusoidal variation of magnitude $100$ $\RR{\mu m}$ on the boundary. Here the differences in the topography are limited to $25$ nm -- well within the working range for precision optics, indicating that the analytical model can be safely used within this range. 

\end{document}